\documentclass[prx,twocolumn,superscriptaddress,showpacs,notitlepage,floatfix,bibnotes,nofootinbib]{revtex4-2}
\usepackage{amsmath,amsthm,amsfonts,amssymb}
\usepackage{graphicx}
\usepackage{tikz}
\usepackage{physics}
\usepackage{dsfont}
\usepackage{float}
\usepackage{xcolor}
\tikzstyle{midarrow} = [decoration={markings, mark= at position 0.6 with {\arrow{Stealth}}}, postaction={decorate}]

\usepackage{graphicx}% Include figure files
\usepackage{physics}
\usepackage{amssymb}

\usepackage{mathtools}

\usepackage{color}

\usepackage{tikz-3dplot}
\usetikzlibrary{3d,shapes,arrows.meta,decorations.pathreplacing,decorations.markings,cd}
\usetikzlibrary{arrows.meta, decorations.markings}

\usetikzlibrary{shapes.geometric, calc}

\usepackage[pdfencoding=auto, psdextra]{hyperref}
\hypersetup{colorlinks=true,citecolor=blue,linkcolor=blue, urlcolor=blue}

\makeatletter
\def\l@subsubsection#1#2{}
\makeatother

\newcommand{\g}{\mathbf{g}}
\newcommand{\h}{\mathbf{h}}
\renewcommand{\k}{\mathbf{k}}
\newcommand{\f}{\mathbf{f}}
\newcommand{\B}{\mathcal{B}}
\newcommand{\C}{\mathcal{C}}

\renewcommand{\Vec}{\mathcal{V}ec}
\newcommand{\Rep}{\text{Rep}}
\newcommand{\wt}{\widetilde}

\newcommand{\nnb}{\nonumber}

\definecolor{FGreen}{RGB}{58,158,92}

\definecolor{lightred}{RGB}{223, 123, 84}
\definecolor{lightbrown}{RGB}{242,216,199}
\definecolor{invisiblegray}{RGB}{238,238,237}
\definecolor{lightgreen}{RGB}{222, 235, 218}
\definecolor{lightcyan}{RGB}{153, 191, 216}
\definecolor{lightcyan2}{RGB}{208, 224, 235}
\definecolor{cyan2}{RGB}{144,201,226}
\definecolor{lightyellow}{RGB}{252, 238, 183}
\definecolor{OrangeRed}{RGB}{223,123,84}
\definecolor{CyanGreen}{RGB}{81,177,184}
\definecolor{fuchsiapurple}{RGB}{180, 0, 255}
\definecolor{Teal}{RGB}{62,138,141}
\definecolor{gold}{RGB}{233,168,58}
\begin{document}

\title{Efficient Preparation of Solvable Anyons with Adaptive Quantum Circuits}
\date{February 8, 2025}

\author{Yuanjie Ren}
\affiliation{Department of Physics, Massachusetts Institute of Technology, Cambridge, Massachusetts 02139, USA}
\author{Nathanan Tantivasadakarn}
\affiliation{Walter Burke Institute for Theoretical Physics and Department of Physics, California Institute of Technology, Pasadena, CA 91125, USA}
\author{Dominic J.~Williamson}
\thanks{Current address: IBM Quantum, IBM Almaden Research Center, San Jose, CA 95120, USA}
\affiliation{School of Physics, University of Sydney, Sydney, New South Wales 2006, Australia}

\begin{abstract}
\noindent
The classification of topological phases of matter is a fundamental challenge in quantum many-body physics, with applications to quantum technology. 
Recently, this classification has been extended to the setting of Adaptive Finite-Depth Local Unitary (AFDLU) circuits which allow global classical communication. 
In this setting, the trivial phase is the collection of all topological states that can be prepared via AFDLU. 
Here, we propose a complete classification of the trivial phase by showing how to prepare all solvable anyon theories that admit a gapped boundary via AFDLU, extending recent results on solvable groups. 
Our construction includes non-Abelian anyons with irrational quantum dimensions, such as Ising anyons, and more general acyclic anyons. 
Specifically, we introduce a sequential gauging procedure, with an AFDLU implementation, to produce a string-net ground state in any topological phase described by a solvable anyon theory with gapped boundary. 
In addition, we introduce a sequential ungauging and regauging procedure, with an AFDLU implementation, to apply string operators of arbitrary length for anyons and symmetry twist defects in solvable anyon theories.  
We apply our procedure to the quantum double of the group $S_3$ and to several examples that are beyond solvable groups, including the doubled Ising theory, the $\mathbb{Z}_3$~Tambara-Yamagami string-net, and doubled $SU(2)_4$ anyons.
\end{abstract}

\maketitle

%\tableofcontents %Switch On/Off for Table of Contents 

\section{Introduction}

Quantum phases of matter have long been a central focus of research in condensed matter physics and quantum information theory. 
These phases have been defined as equivalence classes of ground states of local Hamiltonians on lattice systems with local Hilbert spaces, connected by Finite-Depth Local Unitary (FDLU) circuits~\cite{PhysRevB.82.155138}. 
Of particular interest are systems exhibiting stable long-range entanglement~\cite{Wen_2013}, which in 2+1D are known as Topological Orders (TO)~\cite{Wen:1989iv,RevModPhys.89.041004}. 
The excitations in these systems are called anyons and are described by Modular Tensor Categories (MTCs)~\cite{Kitaev_2006,Lan_2014,Wen_2015}.\footnote{In this work, we use the term anyon theory and modular tensor category (MTC) interchangeably.} 
Topological orders have been the focus of much research interest, due to their applications to fault-tolerant quantum computation~\cite{qdouble,Dennis2001}. 

However, FDLU equivalence is not the only framework for classifying quantum systems. 
Alternative approaches include quantum convolution renormalization~\cite{bu2024magicclassconvolutiongroup}, and augmenting FDLU with measurements and feed-forward operations~\cite{Piroli21,TantivasadakarnmeasureSPT,verresen2022,bravyi2022adaptive,Tantivasadakarn23shortestroute,Tantivasadakarn23Hierarchy,Li23}. 
The latter approach is motivated by the set of operations that can be performed relatively quickly in experiments. 
This leads to an equivalence relation under adaptive finite-depth local unitary (AFDLU) circuits which consist of local unitary gates, single-site measurements, global classical communication, and local unitary feed-forward operations. 
This new equivalence relation connects seemingly distinct topological orders. 
For instance, the quantum doubles of a subnormal series of groups can be created from a product state via finite rounds of FDLU, measurement and feed-forward, which places them in the same AFDLU phase of matter~\cite{Tantivasadakarn23Hierarchy}. 
This framework is expected to distinguish between less complex topological orders including the toric code and more general quantum doubles of solvable groups, which can be prepared via AFDLU, and more complex orders such as the Fibonacci string-net or the quantum double of a non-solvable group, which cannot. 
This classification has direct practical applications in finding anyon states that can be prepared and manipulated efficiently via AFDLU. 
Moreover, near-term quantum devices have finally reached the prerequisites that allow us to efficiently prepare both Abelian and non-Abelian topological phases of matter using adaptive circuits, as has been demonstrated recently~\cite{Iqbal2023,foss2023experimental,iqbal2024non,acharya2024quantum}. 

This work extends AFDLU state preparation to string-net models~\cite{Levin_2005}. 
We demonstrate that any string-net ground state based on an input category that is given by a finite sequence of Abelian group extensions can be prepared via AFDLU. 
We go on to show how to implement string operators of arbitrary length for anyons and symmetry twist defects in solvable anyon theories. 
This implies that all solvable non-Abelian anyon excitations can be prepared and manipulated in constant time via AFDLU, not accounting for classical computation time. 
This class includes examples that are cyclic, i.e. anyons whose fusion with their antiparticle contains themself as an outcome, and examples that have non-integer quantum dimensions. 
We conjecture that our construction is complete, i.e. solvable anyons are the most general class of topological orders that can be prepared via AFDLU in 2+1D. 
If this is not the case, there must be operations beyond gauging and ungauging Abelian symmetries that can be implemented via AFDLU. 

This work is structured as follows. 
In Section~\ref{grading_and_ground_state_preparation} we introduce the concept of graded categories and their application to string-net ground state preparation via AFDLU, using doubled Ising anyons and doubled $SU(2)_4$ anyons as illustrative examples.
In Section~\ref{sect_preparation_of_nilpotent_anyons} we detail the AFDLU implementation of string operators for acyclic, or nilpotent, anyons, which are a simple subclass of solvable anyons. 
In Section~\ref{solvable_anyon_string_op}, we extend this to solvable anyons and introduce our ungauging-regauging approach to implement anyonic string operators via AFDLU. 
This method involves implementing anyon condensation, by ungauging part of the system, to control the anyon fusion process, followed by regauging to restore the original system. We demonstrate this technique on two examples: the quantum double of $S_3$, previously studied in Ref.~\cite{verresen2022}, and the $\mathbb{Z}_3$ Tambara-Yamagami category \cite{TAMBARA1998692,izumi2001,PhysRevB.100.115147,PhysRevB.103.195155}, which features cyclic anyons and showcases our finite-depth preparation method.
In Section~\ref{sec:Discussion} we summarize our results and discuss future directions. 
%-----
In Appendix~\ref{app_StringNet} we review relevant background on string-net models and their symmetry-enriched counterparts. 
In Appendix~\ref{app_G_crossed} we discuss the defect tube algebra for a $G$-crossed modular input theory $\mathcal{C}_G,$ and explain how to find the anyons in the Drinfeld center $\mathcal{Z}(\mathcal{C}_G)$ via gauging. 
In Appendix~\ref{app:CBCircuit} we provide details about implementing controlled plaquette operators on string-net models. 
In Appendix~\ref{app:TY} we give more details about the center of the Tambara-Yamagami category for the group $\mathbb{Z}_3$ and its relation with the $SU(2)_4$ theory. 
%-------------------------
\section{Efficient preparation of solvable string-net ground states}\label{grading_and_ground_state_preparation}
In this section we focus on the problem of preparing topological ground states via adaptive local unitary circuits. 
The AFDLU preparation of states with topological order described by the quantum double of a solvable group $G$ has been the topic of recent research~\cite{TantivasadakarnmeasureSPT,verresen2022,bravyi2022adaptive,Tantivasadakarn23shortestroute,Tantivasadakarn23Hierarchy,Li23}. 
The quantum double model for a group $G$ is equivalent to a string-net model based on the category $\Vec_G$ on a dual lattice~\cite{Levin_2005}. For a solvable group $G$, the category $\Vec_G$ admits a nested series of Abelian gradings. Each Abelian grading corresponds to an Abelian symmetry that is gauged in a sequence to prepare the ground state of the quantum double for the solvable group $G$.

\subsection{Solvable string-nets}

\begin{figure}[htp]
    \centering
    \includegraphics[page=38]{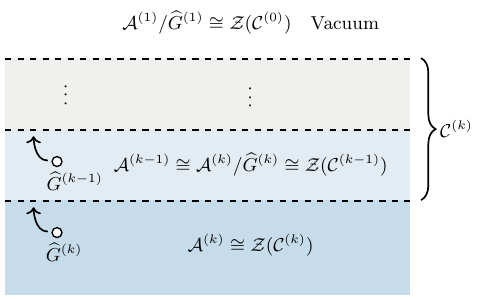}
\caption{
An illustration of a gapped boundary from a solvable anyon theory $\mathcal{A}^{(k)}$ to vacuum that supports pointlike topological defects described by a fusion category $\mathcal{C}^{(k)}$ which is obtained via a sequence of Abelian $G$-extensions, see Eq.~\eqref{eq:grading}.  
The gapped boundary is constructed from a series of Abelian boson condensations described by the groups $\widehat{G}^{(i)}$. 
This leads to the anyon theories $\mathcal{A}^{(i-1)}$ that are found by condensing $\widehat{G}^{(i)}$ bosons in $\mathcal{A}^{(i)}$. 
Dashed lines represent $\widehat{G}^{(i)}$-condensation domain walls. 
The collection of condensed phases can be viewed as a thick gapped boundary to the vacuum with  boundary excitations described by $\mathcal{C}^{(k)}$.
}
\label{fig:GradedBoundary}
\end{figure}

Here, we show how to use a nested Abelian grading structure to prepare a more general class of string-net models, which hosts \textit{solvable anyons}, via AFDLU. 
We refer to Definition~1.2 in Ref.~\cite{Etingof2011weakly} for a formal definition of solvable fusion category.
In this section, we focus on the structure of input categories that lead to solvable anyons via the string-net construction, we defer a direct discussion of solvable emergent anyons to the next section. 
Any solvable anyon theory with a gappable boundary can be realized via a string-net model based on an input fusion category that admits a nested series of Abelian gradings~\cite{gelaki2009centers}, which we describe below. 
A physical picture for such a fusion category is given by pointlike topological defects on a gapped boundary between a solvable anyon theory and the vacuum that is constructed via a sequence of Abelian boson condensations, see Fig~\ref{fig:GradedBoundary}. 

Let $\mathcal{C}^{(0)}=\Vec$ be the trivial fusion category and $G^{(1)}$ a finite group. We denote a $G^{(1)}$-extension of $\mathcal{C}^{(0)}$ by
\begin{align}
    \mathcal{C}^{(1)}=\bigoplus_{\mathbf{g}\in G^{(1)}} \mathcal{C}_{\mathbf{g}}^{(1)},
\end{align} 
where $\mathcal{C}^{(1)}_{\mathbf{1}}=\mathcal{C}^{(0)}$. 
The $G^{(1)}$-grading of the above extension means that fusion respects the group multiplication rule, i.e. the fusion product of $a_{\g}\in \mathcal{C}_g^{(1)}$ and $b_{\h}\in \mathcal{C}_h^{(1)}$ satisfies $a_{\g}b_{\h}=c_{\g \h}\in \mathcal{C}_{\g\h}^{(1)}$.  
We repeat the above process to define a sequence of $G^{(k)}$-extensions such that $\C^{(k)}_{\mathbf{1}}=\C^{(k-1)}$.  
We then have a series of graded categories satisfying 
\begin{align}
\label{eq:grading}
    \Vec\equiv \C^{(0)}\subset \C^{(1)}\subset \cdots \subset \C^{(k)}\equiv \C,
\end{align}
where the gradings are given by ${G^{(0)}\equiv\{\mathbf{1}\}},\, G^{(1)},\cdots,\, G^{(k)}$, respectively. A fusion category $\mathcal C$ with the above property is called a \textit{nilpotent fusion category}, see  Ref.~\cite{Gelaki2008Nilpotent} for a formal definition. 
The smallest $k$ for which the above series exist is called the \textit{nilpotency class} of $\mathcal C$. 
Nilpotent fusion categories have also been called acyclic in the literature~\cite{Dauphinais2016}. 
An equivalent definition of nilpotent fusion categories is that the sequence formed by fusing an anyon with its antiparticle, picking an arbitrary outcome, and then repeating this process, always results in the vacuum after a finite number of steps.

The fusion categories used in this work form a subset of nilpotent fusion categories that satisfy a stricter condition. 
Here, we require that $G^{(i)}$ are all Abelian groups, for $i=1,\ldots,k$. Such categories are called \textit{cyclically-nilpotent fusion categories}\footnote{Note, Eq.~\eqref{eq:grading} does not demand that $\C$ is an Abelian extension of every $\C^{(i)}$ directly. Thus, the only distinction between a sequence of cyclic versus Abelian extensions is the resulting nilpotency class, which translates to the number of rounds of measurements performed in our protocol. Demanding that $\C$ is an Abelian extension of every $\C^{(i)}$ would further restrict $G$ to be supersolvable for a cyclic grading, which we do not require.}~\cite{Etingof2011weakly}.

This definition naturally generalizes the notion of solvability of groups to fusion categories in the following sense: the category $\C=\Vec_G$ is cyclically nilpotent iff $G$ is solvable. Explicitly, let $N^{(i)}$ be the derived series of $G$ defined by
\begin{align}
    N^{(0)} &=G,\\
    N^{(i+1)} &= [N^{(i)},N^{(i)}] \ ;i\ge 1.
\end{align}
Then we have
\begin{align}
    \C^{(i)} &= \Vec_{N^{(k-i)}},\\
    G^{(i)} &= N^{(k-i)}/N^{(k-i+1)}.
\end{align}

For every solvable anyon theory that admits a gapped boundary, a theorem of Ref.~\onlinecite{Etingof2011weakly} states that such an anyon theory is the Drinfeld center $\mathcal Z(\mathcal C)$ of a cyclically nilpotent fusion category $\mathcal C$. Thus, we may realize this anyon theory via the string-net model corresponding to $\mathcal C$. 
See Appendix~\ref{app_StringNet} for a review of string-net models.

We now demonstrate that the ground state of this string-net model can be obtained via a sequence of $k$ steps of Abelian gauging. 
This implies that the string-net ground state can be prepared via an AFDLU with $k$ rounds of measurement and feedforward~\cite{Williamson2020a,TantivasadakarnmeasureSPT}.  
We use $\ket{\psi_{i}}$ to denote the string-net ground state based on the fusion category $\mathcal{C}^{(i)}$. 
Consider the plaquette operators, $B_p^a$, in the string-net model based on $\mathcal{C}^{(i)}$. 
The following sum of plaquette operators forms a representation $\lambda$ of $G^{(i)}$
\begin{align}
    B_p^{\g}:=\frac{1}{\mathcal{D}_i^2} \sum_{a\in \C_{\g}^{(i)}}\ d_aB_p^{a},\quad \g\in G^{(i)},
    \label{eq_B_p_grading_sectors}
\end{align}
i.e. $B_p^{\g}B_p^{\mathbf{h}}=B_p^{\mathbf{gh}}$ and $B_p^{\g^{-1}}=(B_p^{\g})^\dagger$.
This representation can be decomposed into irreducible representations $\lambda=R_1\oplus R_2\oplus\cdots$. 
Since the group $G^{(i)}$ is Abelian, its representations $R$ are equivalent to characters $\chi_R$. 
We define the projector at plaquette $p$ onto a specific representation $R$ among $\oplus_i R_i$ via
\begin{align}
    \Pi^R_p:=\frac{1}{|G|}\sum_{g\in G}\chi^*_R(g)B_p^g.
\end{align}
Here, $\Pi^1_p$ is equal to the plaquette projector of the $\mathcal{C}^{(i)}$ string-net. 

To go from the $\mathcal{C}^{(i-1)}$ string-net to the $\mathcal{C}^{(i)}$ string-net, we first add local degrees of freedom and apply a local unitary circuit that maps to a symmetry-enriched topological (SET) string-net based on $\mathcal{C}^{(i)}$~\cite{PhysRevB.94.235136,cheng2016exactly}. 
We then apply a $G^{(i)}$ gauging map to construct the conventional string-net based on $\mathcal{C}^{(i)}$. 

To construct the symmetry-enriched string-net we take an input $G^{(i)}$-graded fusion category $\mathcal{C}^{(i)}$, viewed as a $G^{(i)}$-extension of $\mathcal{C}^{(i-1)}$. 
The ground state of the $G^{(i)}$ symmetry-enriched string-net is related to the ground state of the conventional string-net based on $\mathcal{C}^{(i-1)}$ via a local unitary, see Appendix~\ref{app_StringNet}.
Let $\ket{\Psi_{(i-1)}}$ denote the ground state of the $\mathcal{C}^{(i-1)}$ string-net, and $|\Psi_{(i-1)}^{\mathrm{SET}}\rangle$ denote the ground state of the $\mathcal{C}^{(i)}$ symmetry-enriched string-net. 
Then we have
\begin{align}
    \label{eq_step_i_set}
    \ket{\Psi_{(i-1)}^{\mathrm{SET}}}=\prod_p CB_p \ket{+}_P\ket{\Psi_{(i-1)}},
\end{align}
where $\ket{+}_P =\bigotimes_{p } \ket{+}_p$ are ancilla states on the plaquettes, with 
\begin{align} 
\ket{+}_p =\frac{1}{\sqrt{|G^{(i)}|}}\sum_{\g\in G^{(i)}}\ket{\g}_p, 
\end{align}
described by the trivial character of $G^{(i)}$, and $CB_p$ are controlled plaquette operators defined by
\begin{align}
    CB_p\ket{\g }_p\ket{\Psi}=\ket{\g}_p B_p^{\g}\ket{\Psi}.
    \label{eq_def_controlled_plaquette}
\end{align}
See Appendix~\ref{app_StringNet} for further discussion of the map between the conventional string-net for $\mathcal{C}^{(i-1)}$ and the symmetry-enriched string-net for $\mathcal{C}^{(i)}$.

The symmetry-enriched string-net state $|\Psi_{(i-1)}^{\mathrm{SET}}\rangle$ transforms under the on-site global $G^{(i)}$ symmetry $\prod_{p} L_p(\g)$, where $L(\g)$ denotes left multiplication by $\g\in G^{(i)} $. 
The $G^{(i)}$ symmetry-enrichment of $\mathcal{Z}(\mathcal{C}^{(i-1)})$ topological order in this string-net is described by the relative center  $\mathcal{Z}_{G^{(i)}}(\mathcal{C}^{(i)}_{G^{(i)}})$, see Ref.~\cite{williamson2017}. 

The ground state of the string-net with input category $\mathcal{C}^{(i)}$ is obtained by gauging the $G^{(i)}$-symmetry of $|\Psi_{(i-1)}^{\mathrm{SET}}\rangle$. 
In this case, gauging is equivalent to simply projecting each plaquette state onto $\bra{+}_p$~\cite{williamson2017}. 
This results in the gauged string-net state
\begin{align}
    \ket{\Psi_{(i)}}=\bra{+}_P \prod_p CB_p\ket{+}_P \ket{\Psi_i} .
    \label{eq_step_i_gauging}
\end{align}
This is simply the ground state of the $\mathcal{C}^{(i)}$ string-net since $\bra{+}_p \prod_p CB_p\ket{+}_p$ implements the plaquette projector $\Pi^1_p$, which matches the string-net plaquette projector for $\mathcal{C}^{(i)}$. 
The vertex constraints of the $\mathcal{C}^{(i)}$ string-net are already satisfied by the symmetry-enriched string-net state, see Appendix~\ref{app_StringNet}, and they remain satisfied, as they commute with $\Pi^1_p$.

Importantly, the structure of the Abelian grading allows the projection $\bra{+}_P=\prod_{p\in P}\bra{+}$ to be implemented by measurement and feedforward~\cite{Williamson2020a,TantivasadakarnmeasureSPT}. 
For example, suppose the measured state corresponds to an excitation 
\begin{align}
\ket{\chi}_p = \frac{1}{|G^{(i)}|} \sum_{\mathbf{g} \in G^{(i)}} \chi^*(\mathbf{g}) \ket{\mathbf{g}} ,\label{eq_character_p}
\end{align}
described by some character $\chi$ of $G^{(i)}$.
This results in the plaquette projector $\Pi_p^{\chi}$ on the post-measurement state. 
To remove such excitations we use generalized character operators on edges that are defined by  $\widetilde{\chi}_e\ket{a_{\mathbf{g}}} = \chi(\g) \ket{a_{\mathbf{g}}}$ where $a_{\g} \in \mathcal{C}^{(i)}_{\g}$. 
These character operators satisfy $\widetilde{\chi}_e \Pi_p^{\chi}=\Pi_p^{1} \widetilde{\chi}_e$ for $e \in p$ with a matching orientation. 
When applied to a $\mathcal{C}^{(i)}$ string-net state, $\widetilde{\chi}_e$ creates a particle-antiparticle pair of $\chi^*_p$, $\chi_{p'}$, bosons on adjacent plaquettes $p,p'$, with matching and opposing orientations, respectively. 
Hence, we can use character operators to pair up a $\ket{\chi}_p$ measurement outcome with another plaquette measurement $\ket{\chi^*}_{p'}$ at $p'$ via a string operator $\prod_{e\in \gamma} \widetilde{\chi}_e$ acting on the edge qudits along a path $\gamma$ in the dual lattice that satisfies $(\partial \gamma)_0 =p$ and $(\partial \gamma)_1=p'$ to annihilate the pair of excitations, leaving $\ket{+}$ states on the plaquettes. 
More generally, we can move all plaquette charge excitations that result from measurement to a single location via a product of $\prod_{e\in \gamma} \widetilde{\chi}_e$ string operators. 
The fusion of all plaquette charges must result in the trivial charge as the original state is $G^{(i)}$ symmetric. 
This process is similar to pairing up syndromes when initializing the toric code, and as in that case, the choice of correction operator does not matter as all choices result in the same initialized state. 

In the following sections, we demonstrate a number of examples that can be prepared following the above procedure: the Ising category, the quantum double of $S_3$, the $\mathbb{Z}_3$ Tambara-Yamagami category, and doubled $SU(2)_4$.

%-----------------
\subsection{Example: doubled Ising anyon theory}\label{subsection_DIsing}
The doubled Ising anyon theory $\mathcal{Z}(\text{Ising})$ is nilpotent, and hence we can use the procedure introduced above to prepare the ground state of this model. 
The Ising category can be obtained from the trivial theory via a pair of $\mathbb{Z}_2$ extensions
\begin{align}
\underbrace{\mathbf{1}}_{\Vec} \xrightarrow{\mathbb{Z}_2}  \underbrace{1\oplus \psi}_{\Vec_{\mathbb{Z}_2}} \xrightarrow{\mathbb{Z}_2}\underbrace{ (1\oplus \psi) \oplus \sigma}_\text{Ising} ,\label{eq_Ising_extensions}
\end{align}
resulting in the following fusion rules
\begin{align}
    \psi\otimes \psi=\mathbf{1},\quad\psi \otimes \sigma=\sigma,\quad \sigma\otimes \sigma=\mathbf{1}\oplus \psi.
\end{align}
The centers of the above categories correspond to emergent anyons appearing in a sequence of $\mathbb{Z}_2$ gaugings 
\begin{align}
\Vec \xrightarrow{\text{Gauge } \mathbb{Z}_2}  \underbrace{\mathcal Z(\Vec_{\mathbb Z_2})}_{\mathrm{TC}}  \xrightarrow{\text{Gauge } \mathbb{Z}_2} \underbrace{\mathcal Z(\text{Ising})}_{\text{Ising} \times \overline{\text{Ising}}} ,
\end{align}
where TC denotes the toric code anyons.

We use the following map $\ket{\mathbf{1}}\to \ket{0}$, $\ket{\psi}\to \ket{1}$, ${\ket{\sigma}\to \ket{2}}$ to represent string types of $\Vec_{\mathbb{Z}_2}$ and Ising in the computational basis of a qubit, and qutrit, respectively.  

\subsubsection{The first gauging}
Let $E,P$ denote the set of all edges and all plaquettes of the lattice.
We initialize a state
\begin{align}
    \ket{\Psi_0}_P=\ket{+}_P=\bigotimes_{p\in P} \ket{+}_p
\end{align}
where 
\begin{align}
   \ket{\pm}_p=(\ket{\mathbf{1}}_p\pm \ket{\psi}_p)/\sqrt{2}.
\end{align}  
We then apply the 2+1D Kramers-Wannier~\cite{Kramers1941} (KW) map to obtain the ground state of the toric code,
\begin{align}
KW_{EP}^{\mathbb{Z}_2}=&\bra{+}_P \left(\prod_p \prod_{e\in \partial p} CX_{p\to e}\right)\ket{0}_E\\
\ket{\Omega}_{\mathbb{Z}_2}=&KW_{EP}^{\mathbb{Z}_2}\ket{\Psi_0}_P, \label{eq_toric_code_ground_state}
\end{align}
where $\prod_{e\in \partial p}CX_{p\to e}$ is the controlled plaquette operator that appeared in Eq.~\eqref{eq_step_i_gauging}.
Each $\bra{\pm}_p$ measurement corresponds to the plaquette operator taking the value 
\begin{align}
    B^\psi_p=\prod_{e\in \partial p}X_e=\pm 1
\end{align}
on plaquette $p$ of the string-net lattice. 
Equivalently, $B_p=(B^{\mathbf{1}}_p+B^{\psi}_p)/2$ takes the value $1$ or $0$, respectively. 
These measurement outcomes correspond to the electric excitations of the toric code, which can be removed by pairing them up in finite depth using Pauli $Z$ strings  related to  the first $\mathbb{Z}_2$ grading in Eq.~\eqref{eq_Ising_extensions}. 

%----
\subsubsection{The second gauging}\label{sect_the_second_gauging}
We now embed each qubit $\{\mathbf{1},\psi\}$ into a qutrit $\{\mathbf{1},\psi,\sigma\}$ where $\sigma$ carries the nontrivial grading of the second $\mathbb{Z}_2$ group. 
Hence the toric code ground state in the previous step can be viewed as the ground state of the following Hamiltonian 
\begin{align}
    H=-\sum_v Q_v -\sum_p B_p -\sum_e P_e
\end{align}
where $P_e$ is the projector onto the $\{\mathbf{1},\psi\}$ subspace, such that the $\ket{2}\equiv \ket{\sigma}$ degrees of freedom do not appear in the string-net basis state expansion of states in the ground state subspace.
In the gauging process, the vertex operator which enforces the fusion rules  is always satisfied, $Q_v=+1$. 
After gauging the edge degrees of freedom $\ket{\sigma}$ are fluctuated and become unfrozen. 

Using the $\mathbb{Z}_2$ grading of Ising, we define plaquette operators
\begin{align}
    B^0_p:=\frac{1}{2}(B_p^{\mathbf{1}}+B_p^\psi),
    \quad B_p^1:=\frac{\sqrt{2}}{2}B_p^\sigma.
\end{align}
These operators form a representation of $\mathbb{Z}_2$
\begin{align}
  \left(B_p^1\right)^2=B_p^0,
\end{align}
as explained in Eq.~\eqref{eq_B_p_grading_sectors}. 
Using the general formula in Eq.~\eqref{eq_def_controlled_plaquette} for $G=\mathbb{Z}_2$
\begin{align}
   (CB)_p\ket{i}_p\ket{\psi}_E:=B_p^i\ket{i}_p\ket{\psi}_E,\quad i=0,1. 
\end{align}
Now we use controlled-$B^\sigma_p$ gates to prepare a state with an appropriate $\mathbb{Z}_2$ symmetry action to gauge. 
Let $\ket{+}_P$ represent the plaquette state for the second $\mathbb{Z}_2$ grading in Eq.~\eqref{eq_Ising_extensions}.
At this point the plaquette ancilla qubits from the first gauging procedure have been measured out and discarded. 
We then have
\begin{align}
KW_\text{Ising}^{\mathbb{Z}_2}:=&\bra{+}_P \prod_p CB_p \ket{\Psi}_E\\
\ket{\Omega}_\text{Ising}=&KW_\text{Ising}^{\mathbb{Z}_2}\ket{+}_P,
\end{align}
where $\ket{\Psi}_E=\ket{\Omega}_{\mathbb{Z}_2}$ is the wavefunction on the edge system $E$ that is fed into the generalized KW duality.
In the example here, this is the ground state of the $\mathbb{Z}_2$ topological order we have obtained from the previous gauging procedure, see Eq.~\eqref{eq_toric_code_ground_state}. 
See Appendix~\ref{app:CBCircuit} for an explicit circuit description of the $CB_p$ operators. 

The two possible measurement outcomes $\bra{\pm}_p$ at plaquette $p$ correspond to projectors for the following two anyonic excitations in Double Ising
\begin{align}
    \Pi_{vac}=&\frac{1}{4}(\mathds{1}+B_p^\psi+\sqrt{2}B_p^\sigma)\label{eq_Pi_vac}\\
    \Pi_{\psi\overline{\psi}}=&\frac{1}{4}(\mathds{1}+B_p^\psi- \sqrt{2}B_p^\sigma).\label{eq_Pi_psi_psibar}
\end{align}
Similar to above, the $\bra{-}_p$ outcomes can be paired up and fused to the vacuum. 
Using the second $\mathbb{Z}_2$ grading in Eq.~\eqref{eq_Ising_extensions}, we define the corresponding generalized character operator $\widetilde{\chi}=\text{Diag}(1,1,-1)$.
Hence, $\widetilde\chi$ assigns the unique element from $\mathcal{C}_\sigma=\{\sigma\}$ to $-1$ and assigns $\mathcal{C}_{\mathbf{1}}=\{\mathbf{1},\psi\}$ to $+1$.
One can apply a string of $\widetilde\chi$ operators to clean up the undesired  $\bra{-}_p$ measurement outcomes, which correspond to the application of projectors $\Pi_{\psi \bar{\psi}}$ from Eq.~\ref{eq_Pi_psi_psibar}.
%-----
To summarize, the the explicit circuit to prepare the ground state of the doubled Ising string-net model is given by
\begin{align}
 \bra{+}_P \prod_p CB_p \ket{+}_P\bra{+}_P \prod_p \prod_{e\in \partial p} CX_{p\rightarrow e}\ket{0}_E \ket{+}_P
\end{align}
which can be implemented by a constant-depth adaptive circuit, requiring two rounds of measurement and feedforward. 
This constitutes the only constant-depth adaptive circuit for the preparation of the doubled Ising topological order that has appeared in the literature to date.

%-----------
\subsection{Example: Doubled $SU(2)_4$}
In the previous example, the input fusion category was cyclically nilpotent. Here, we briefly describe how to prepare string-nets corresponding to solvable but not cyclically nilpotent fusion categories.

The idea of the construction is as follows. 
The theorem of Ref.~\onlinecite{Etingof2011weakly} guarantees that any solvable fusion category $\mathcal C$ is Morita equivalent to a cyclically nilpotent fusion category $\mathcal C'$. 
In physics terminology, this means that even if $\mathcal C$ is not cyclically nilpotent, there exists a different cyclically nilpotent fusion category $\mathcal C'$ for which the corresponding string-net ground state exhibits the same topological order. Thus, we may first prepare the string-net corresponding to $\mathcal C'$ using measurements and feedforward. Then, we may use a constant depth circuit to map the string-net ground state of $\mathcal C'$ back to the string-net ground state of $\mathcal C$. An explicit circuit between the fixed point string-nets states is given in Ref.~\cite{Lootens22Morita}, and more generally, such a circuit exists even away from the fixed point~\cite{KimRanard24}.

As an example, let us consider $\mathcal C=SU(2)_4 \equiv \{0,\frac{1}{2}, 1,\frac{3}{2}, 2 \}$, which is not a cyclically nilpotent fusion category. This is because it contains fusion rules ${1 \otimes 1 = 0 \oplus 1 \oplus 2}$ and therefore cannot admit a grading. We can instead choose a different input category $\mathcal C' = \mathcal C_{S_3}$ which has the same bulk topological order~\cite{Frohlich04,Vanhove_2022}.

The fusion category $ \mathcal C_{S_3} = \Vec_{S_3} \oplus \sigma \oplus \sigma'$, where $\Vec_{S_3} = 1 \oplus r \oplus r^2 \oplus s \oplus sr \oplus sr^2$ is labeled by elements of the group $S_3$. The fusion rules of $\Vec_{S_3}$ follow the multiplication rules of $S_3 = \langle r,s|r^3 = s^2 = (sr)^2 =1 \rangle$. The remaining fusion rules are
\begin{align}
\sigma \otimes \sigma = \sigma' \otimes \sigma' &= 1 \oplus r \oplus r^2, \\
\sigma \otimes \sigma' =\sigma' \otimes \sigma &= s \oplus sr \oplus sr^2,\\
\sigma \otimes r^\alpha = \sigma' \otimes s r^\alpha  &= \sigma, \\
\sigma \otimes sr^\alpha = \sigma' \otimes r^\alpha &= \sigma',
\end{align}
for $\alpha=0,1,2$. Thus, we see that $ \mathcal C_{S_3} $ admits a $\mathbb Z_2 \times \mathbb Z_2$ grading 
\begin{align}
(\mathcal C_{S_3})_{00} &= 1 \oplus r \oplus r^2 \cong \Vec(\mathbb Z_3)\\
 (\mathcal C_{S_3})_{10} &= s \oplus sr \oplus sr^2\\
(\mathcal C_{S_3})_{01} &= \sigma\\
(\mathcal C_{S_3})_{11} &= \sigma'   
\end{align}
and is therefore nilpotent. 
We label the first and second $\mathbb Z_2$ extensions as charge conjugation $\mathbb Z_2^C$ and electromagnetic duality $\mathbb Z_2^{em}$, respectively. The $F$-symbols of this fusion category can be found in Ref.~\onlinecite{Vanhove_2022}.

\begin{widetext}
The corresponding extensions are summarized below. 
\begin{equation}
  \begin{tikzcd}
&&\Vec_{S_3} \arrow[dr, "\mathbb{Z}_2^{em} "]&\\
\underbrace{\mathbf{1}}_{\Vec} \arrow[r, "\mathbb{Z}_3 "]&\underbrace{1 \oplus r \oplus r^2}_{\Vec_{\mathbb Z_3}} \arrow[ur, "\mathbb{Z}_2^C "]  \arrow[dr,swap, "\mathbb{Z}_2^{em} "] \arrow[rr, "\mathbb{Z}_2^{C} \times \mathbb{Z}_2^{em} "] && \mathcal C_{S_3}\\
&&TY(\mathbb Z_3)  \arrow[ur,swap, "\mathbb{Z}_2^C "]&
\end{tikzcd}  
\end{equation}

The sequence of emergent anyon theories that are obtained by taking the centers of the above categories are related by gauging, as shown below. 
\begin{equation}
\begin{tikzcd}
&&\mathcal{D}(S_3) \arrow[dr, "\text{Gauge }\mathbb{Z}_2^{em} "]&\\
\Vec\arrow[r, "\text{Gauge }\mathbb{Z}_3 "]&\underbrace{\mathcal{D}(\mathbb Z_3)}_{\mathbb Z_3^{(1)} \boxtimes \mathbb Z_3^{(-1)}} \arrow[ur, "\text{Gauge }\mathbb{Z}_2^C "]  \arrow[dr, swap,"\text{Gauge }\mathbb{Z}_2^{em} "] \arrow[rr, "\text{Gauge }\mathbb{Z}_2^C \times \mathbb{Z}_2^{em} "] && \underbrace{\mathcal{Z}(\mathcal C_{S_3})}_{SU(2)_4 \boxtimes \overline{SU(2)_4}}\\
&&\underbrace{\mathcal Z(TY(\mathbb Z_3))}_{\mathbb Z_3^{(1)} \boxtimes \overline{SU(2)_4}}  \arrow[ur,swap, "\text{Gauge }\mathbb{Z}_2^{C} "]&
\end{tikzcd}
\end{equation}
\end{widetext}
This implies that we may prepare a state supporting doubled $SU(2)_4$ anyons with either three rounds of gauging or just two rounds of gauging, by combining the last two steps into a single round. 
Therefore, the doubled $SU(2)_4$ topological order can be prepared with two rounds of measurement. 
We defer the details of deriving this gauging sequence to Appendix~\ref{app:TY}.

%------------
\section{Efficient string operators for solvable anyons}
\label{sec:StringOperators}

In this section we first describe a high level strategy for AFDLU implementation of string operators for nilpotent anyons. 
We then describe a generalization of this strategy for AFDLU implementation of string operators for solvable anyons. 
Our strategy also covers twist defects in symmetry-enriched solvable anyon theories. 
We demonstrate our approach via several lattice model examples. 

\subsection{Nilpotent anyon string operators}
\label{sect_preparation_of_nilpotent_anyons}

\begin{figure*}[t]
    \centering
    \includegraphics[page=7]{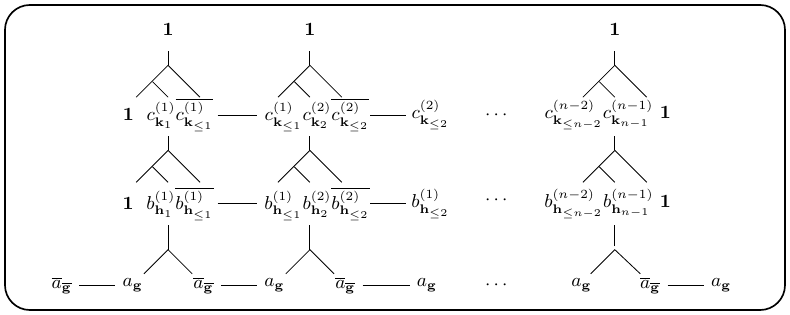}
%--------------------
\caption{Sequential fusion implementation of a string operator to prepare a pair of nilpotent anyons $\overline{a}-a$. The illustrated theory admits a sequence of graded subcategories $\B^{(0)}\subset \cdots \subset \B^{(3)}$. For simplicity we have illustrated the simple case where $b_{\h_{j\leq n-1}}^{(n-1)}=\mathbf{1}$. }
\label{string_of_nilpotent_anyons}
\end{figure*}

%---------
A simple physical description of a nilpotent anyon theory is one where any sequence generated by fusing an anyon with its antiparticle, measuring the outcome anyon type, and then repeating the process terminates with certainty after a constant number of steps. 

In this section we consider nilpotent braided anyon theories. Nilpotent fusion categories are defined in Eq.~\eqref{eq:grading}. While nilpotent fusion categories can generally have non-Abelian grading groups $G_i$, in a nilpotent braided fusion category (and hence for nilpotent anyon theories) the fusion order can be exchanged, and hence $G_i$ must be Abelian. 

To avoid confusion, we reserve the symbol $\mathcal{C}$ for the input fusion category to a string-net construction. 
We use $\mathcal{B}$ to denote a nilpotent anyon theory, and denote the intermediate categories in the grading series by $\mathcal{B}^{(j)}$.   Namely, we consider the nilpotent anyon theory $\mathcal{B}$, which, by definition, admits a sequentially-graded structure
\begin{align} 
\Vec\equiv \B^{(0)}\subset \B^{(1)}\subset \cdots \subset \B^{(k)}\equiv \mathcal{B},\label{eq_a_sequence_of_graded_categories}
\end{align}
with grading groups $(G^{(i)})_{i=0}^k$. 
Here, we note that while $\B^{(i)}$ are braided fusion categories, they do not have to be modular. 
We also note that $\mathcal{B}$ may be chiral. 
To move an anyon over a distance $n$, or to create a particle-antiparticle pair, $\overline{a}_{\overline{\g}}-a_{\g}$, separated by a distance $n$, we first create $O(n)$ particle-antiparticle pairs $\overline{a}_{\overline{\g}}-a_{\g}$ separated by a constant length, which we fix to 1, as illustrated in Figure~\ref{string_of_nilpotent_anyons}. 
Here $\overline{\g}$ denotes the inverse group element $\g^{-1}$. 
We then simultaneously fuse $a_{\g}$ from the $\ell$-th pair with $\overline{a}_{\overline{\g}}$ from the $(\ell+1)$-th pair and measure the resulting anyon type. 
Let $b^{(\ell)}_{\h_\ell}$ denote the measured fusion result. 
Due to the nilpotent structure, we have ${b^{(\ell)}_{\h_\ell}\in \mathcal{B}^{(k-1)}}$, which is the trivial $G^{(k)}$-sector $\mathcal{B}^{(k)}_{\mathbf{1}}=\mathcal{B}^{(k-1)}$.
Next, we define $\h_{\leq j}:=\h_1\h_2\cdots \h_j$, where $\h_i\in G^{(k-1)}$ are the grading labels of the measured anyons in $\mathcal{B}^{(k-1)}$. 
We  create pairs of anyons $\overline{b}^{(j)}_{\h\leq (j-1)}-b^{(j)}_{\h\leq j}$ such that  
\begin{align}
    b_{\h_{\leq j-1}}^{(j-1)} \otimes b_{\h_j}^{(j)}\otimes \overline{b^{(j)}_{\h_{\leq j}}} \in \mathcal{B}_{\mathbf{1}}^{(k-1)}
    ,
\end{align}
where $\otimes$ denotes the fusion product. 
In particular, $\overline{b_{\h_{j\leq 0}}^{(0)}}=\mathbf{1}$ by construction. 
In the simplest case, $b_{\h_{j\leq n-1}}^{(n-1)}=\mathbf{1}$. 
If this is not the case, we can simply fuse it into the dangling $a_{\g}$ anyon at the far end of the string of anyons. 
The same process is repeated $(k-1)$ times until we reach the trivial category $\B^{(0)}\equiv \Vec$. 
At this point we have created a particle-antiparticle pair of anyons $\overline{a}-a$ separated by a distance $n$. 
The anyon at the far end of the string operator must result in $a$, even if it is fused with other anyons at intermediate steps. 
This is because the $\overline{a}$ anyon at the other end of the string operator is not altered during the process of applying the string operator, and the global charge of the region containing the final pair of anyons must be neutral. 

To turn the above procedure into an AFDLU operator on the lattice, we rely on the existence of local unitary gates to apply short string operators, fuse anyons, and measure anyon type, see Ref~\cite{Liu2021} for string-net models. 
We remark that the first step of the above process can be applied directly to twist defects of potentially non-Abelian symmetry groups. 
A similar procedure has previously been applied to implement dualities of 1+1D quantum spin chains, see Ref.~\cite{Lootens2023LDU}.

%---------------

\subsection{Solvable anyon string operators}\label{solvable_anyon_string_op}
%------------------
In this section, we consider solvable anyon theories. 
First, we focus on solvable anyon theories that admit gapped boundaries to the vacuum. 
Such anyon theories can be realized as the Drinfeld center of some cyclically nilpotent input category $\C$, admitting a grading sequence $(\C^{(i)})_{i=0}^{k}$ as in Eq.~\eqref{eq:grading}. 
The grading sequence structure of the input category has a consequence for the resulting Drinfeld center.
In particular, since $(G^{(i)})_{i=0}^k$ is a sequence of Abelian groups, $\mathcal Z(\mathcal{C}^{(i)})$ is obtained from $\mathcal Z(\mathcal{C}^{(i-1)})$ by extending it to a $G^{(i)}$-crossed braided fusion category followed by gauging $G^{(i)}$ ($G^{(i)}$-equivariantization).
In terms of physical operations, a solvable anyon theory has the following property
\begin{align}
    \mathcal{Z}(\C^{(i-1)})\xrightleftharpoons[\text{Condense $Rep(G^{(i)})$}]{\text{Gauge $G^{(i)}$}} \mathcal{Z}(\C^{(i)}) ,\label{eq_gauging_condensing}
\end{align}
See Refs.~\cite{PhysRevB.100.115147,williamson2017}.
Importantly, for Abelian symmetries the emergent anyons $a^{(i)}\in\mathcal{Z}( \C^{(i)})$ inherit a $G^{(i)}$-grading since $\mathcal{Z}(\C^{(i)})$ is obtained from the $G^{(i)}$-crossed extension of $\mathcal{Z}(\C^{(i-1)})$ via gauging $G^{(i)}$. This gauging operation maps the set of $\mathrm{g}$-defects to $\mathrm{g}$-dyons in the gauged theory. 
For example, $\mathcal{Z}(S_3)$ can be obtained by gauging the $\mathbb{Z}_2$-charge conjugation symmetry of $\mathcal{Z}(\mathbb{Z}_3)$. The emergent anyons (see Table~\ref{table_of_S3_anyons}) inherit a grading from this $\mathbb{Z}_2$  symmetry: anyons $\widetilde{D}$ and $\widetilde{E}$ are graded by $1\in \mathbb{Z}_2$, while all other anyons are graded by $0\in\mathbb{Z}_2$. See Ref.~\cite{Beigi_2011} for the full data.

\begin{figure*}[t]
    \centering
    \includegraphics[page=8]{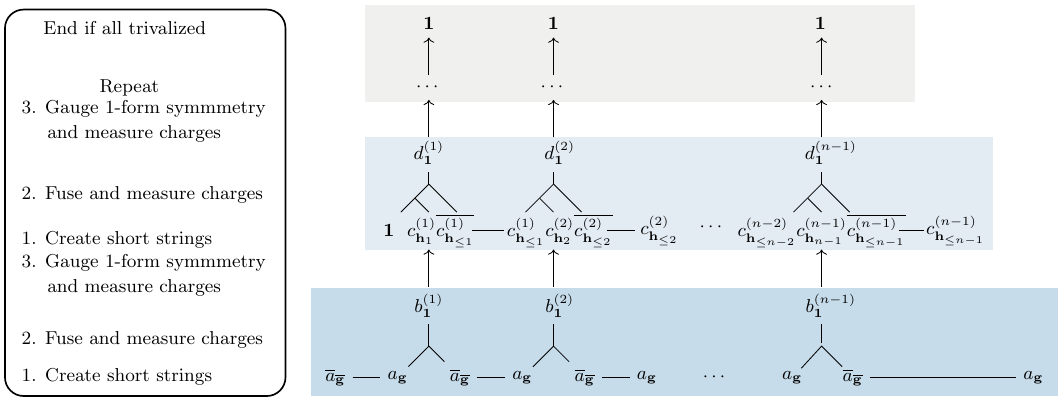}
%--------------------
\caption{Sequential fusion and gauging implementation of a string operator to create a pair of solvable anyons. Note that if $a$ and $\overline{a}$ are cyclic, one may skip the first fusion process to $b_{\mathbf{1}}^{(j)}$, and choose to condense  $a_{\g}$ and $\overline{a}_{\overline{\g}}$ at the beginning. Depending on the solvability of the fusion result one may choose to condense anyons in a neighborhood of the string operator and repeat the fusion process until all $\mathbf{1}$ outcomes are reached. One then gauges the region containing the string operator back to the original topological order $\mathcal{Z}(\mathcal{C})$.}
\label{string_of_solvable_anyons}
\end{figure*}
We now discuss why the procedure to apply nilpotent anyon string operators fails for solvable anyons. Consider applying the procedure for nilpotent anyons. 
We first applying $n$ pairs $\overline{a}_{\overline{\g}}-a_{\g}$ of length 1, and fuse $\overline{a}_{\overline{\g}}$ from the $j$-th pair with $a_{\g}$ from the $(j+1)$-th. 
Again, the fusion result $b_{\mathbf{1}}^{(j)}$ may not be trivial, but it must lie in the trivial sector $\mathcal{C}^{(k)}_{\mathbf{1}}$. 
A new feature for solvable anyons is that $b_{\mathbf{1}}^{(j)}$ can be cyclic, meaning the fusion of $b_{\mathbf{1}}^{(j)}$ with its antiparticle can result in  $b_{\mathbf{1}}^{(j)}$ itself. 
For example, the non-Abelian gauge charge anyon $\widetilde{C}$ in $\mathcal{D}(S_3)$ is cyclic (see Table.~\ref{table_of_S3_anyons}). 
A successful fusion result to remove a pair of such anyons is $\widetilde{C}\otimes \widetilde{C}\to \mathbf{1}\oplus \widetilde{B}$. 
If we suppose these outcomes occur with probability $2/3$, then after depth $D$, the success rate of the string operator growing by $1$ is $1-(2/3)^D$. 
We conclude that the depth of a probabilistic application of a string operator for $\widetilde{C}$ of length $n$ is ${D=O( n\log(1/\epsilon))}$, for an error tolerance $\epsilon$.

We now introduce an alternative procedure to fuse cyclic anyons and reach the vacuum deterministically in finite depth for solvable anyon theories. 
To achieve this, we condense a bosonic $\mathrm{Rep}(G^{(k)})$ subtheory of $\mathcal{C}^{(k)}$ in a neighborhood of the string operator to produce $\mathcal{Z}(\C^{(k-1)})$ anyons. 
This condensation is implemented controllably by gauging a $\mathrm{Rep}(G^{(k)})$ 1-form symmetry generated by the string operators of these Abelian bosons. 
This is also known as ungauging a global $G^{(k)}$ symmetry. 
Therefore, as shown in Figure~\ref{string_of_solvable_anyons}, we implement anyon condensation within a neighborhood of the string operator such that $b_{\mathbf{1}}^{(1)},\cdots b_{\mathbf{1}}^{(n-1)}$ are mapped to anyons in $\mathcal{Z}(\mathcal{C}^{(k-1)})$, while anyons outside of this region are unaffected. 
In particular, the anyons $\overline{a}_{\overline{\g}}$ and $a_{\g}$ at the endpoints of the string operator are unchanged. 
Let $c_{\h_j}^{(j)}$, with  $\h_j\in G^{(k-1)}$, be the excitation resulting from $b^{j}_{\mathbf{1}}$ after the condensation process.
In general, anyons from $\mathcal{Z}(\mathcal{C}^{(k)})$ can split during condensation, resulting in a superposition of different anyons in $\mathcal{Z}(\mathcal{C}^{(k-1)})$. For example, depending on the internal state of the anyon $\widetilde{C}$ of $\mathcal{Z}(S_3)$ it can split into a superposition of $e$ and $e^*$ anyons in $\mathcal{Z}(\mathbb{Z}_3)$. 
Hence, an anyonic charge measurement is required to specify each charge $c_{\h_j}^{(j)}$ in the condensed region.  
We again define $\h_{\leq j}:=\h_1\h_2\cdots \h_j$, and create pairs $\overline{c_{\h_{\leq j}}^{(j)}}-c_{\h_{\leq j}}^{(j)}$ such that the following fusion product lies in the trivial sector
\begin{align}
   c_{\h_{\leq j-1}}^{(j-1)}\otimes c_{\h_j}^{(j)}\otimes \overline{c_{\h_{\leq j}}^{(j)}} \in \mathcal{Z}(\mathcal{C}^{(k-1)})_{\mathbf{1}}. \label{eq_c_h_j}
\end{align} 
Let $d_{\mathbf{1}}^{(j)}$ be the measured result of the above fusion. 
We can then repeat the above procedure by further gauging $\textrm{Rep}(G^{(k-1)})$ 1-form symmetry and measuring the resulting charges. 
For any solvable anyon theory, after several repetitions we are guaranteed to find only trivial anyons $\mathbf{1}$ along the string operator with $\overline{a}-a$ at the end points. 
We then sequentially gauge global $G^{(i)}$ symmetries on the condensed neighborhood of the string operator to restore the original topological order.

The above procedure is similar to Sec.~\ref{sect_preparation_of_nilpotent_anyons}. 
One point of difference is that for solvable anyons, we cannot simply assume $c_{\h_{\leq n-1}}^{(n-1)}$ is the trivial anyon $\mathbf{1}$. 
This is because we cannot directly fuse the residual charge at each step into the $a$ anyon at the end of the string, as that would require tunnelling $c_{\h_{\leq n-1}}^{(n-1)}$ through a domain wall at the boundary of the condensed region.
Instead, we follow another strategy where each residual anyon is left near the boundary of the condensed region, and outside the following condensations.
The set of residual anyons remains until the end of the procedure, when all regions are gauged to restore the original topological order. 
The gauged residual anyons are then fused into the $a$ anyon at the end of the string operator. 
Following similar logic to the nilpotent case, after these fusions, the result must be $a$ due to a global charge neutrality constraint. 

General solvable anyon theories can be chiral and are constructed by a sequence of Abelian $G$-extensions and $G$-equivariantizations. 
This can include a combination of braided $G$-extensions, without gauging $G$, and $G'$-crossed braided extensions that are followed by gauging $G'$. 
To implement string operators for general solvable anyon theories, one can combine the strategies applied in this section for solvable anyons with a gapped boundary and the previous section for nilpotent anyons. 
This strategy suffices to implement symmetry twist defects in symmetry-enriched solvable anyon theories, as such defects deterministically fuse to the anyon sector, even for non-Abelian groups. 
Similar to the above section, the procedure described in this section leads to explicit AFDLU string operators given local unitaries that implement anyon pair creation~\cite{Liu2021}, measurements that implement fusion and measure anyonic charge, and AFDLUs that implement gauging of Abelian global and 1-form symmetries. 

The recipe for implementing solvable anyon string operators in this section can be applied to prepare ground states of solvable anyon theories via AFDLU. 
This proceeds by measuring into a random configuration of solvable anyons and using the above string operators to fuse these anyons to the vacuum in a finite number of steps. 

%-------
\subsubsection{Ungauging and regauging a global symmetry on a simply connected region}
Our approach to implementing a solvable anyon string operator via AFDLU relies on ungauging and subsequently regauging an Abelian global symmetry on the simply connected region that contains the string operator. 
Here, we show that this process can be performed via AFDLU without creating any undesired defects. 

The ungauging step is equivalent to gauging a 1-form symmetry generated by the Abelian gauge charges obtained by gauging an Abelian 0-form global symmetry. 
We consider 1-form symmetries that act on edges associated with closed cycles in the cycle group $Z_1(C,\hat{G})$ on a two-dimensional cell complex $C$ with coefficients in the dual of an Abelian group $\hat{G}$, which is isomorphic to $G$. 
Each cycle is represented by an on-site 1-form symmetry operator 
\begin{align}
    U({z})= \prod_{e\in C} U_e({z_e}) ,
\end{align}
for $z=\sum_{e\in C}z_e \cdot e\in Z_1(C,\hat{G})$, $z_e\in \hat{G}$, and where $U({\cdot})$ denotes a representation of a cycle, and $U_e(\cdot)$ is a representation of a dual group element on a single edge.

For the gauging procedure, we focus on the subgroup of cocycles $Z^1(R,G)$ that are supported within a simply connected subregion $R$ of the cellulation. 
To gauge this 1-form subgroup we introduce $\mathbb{C}[G]$ gauge fields on the vertices initialized in the state 
\begin{align}
    \ket{+}=\frac{1}{|G|^\frac{1}{2}}\sum_{\g\in G} \ket{\g}
\end{align} 
and measure the set of 1-form Gauss's law projectors on each edge
\begin{align}
   \Pi_e(\g) := \frac{1}{|\hat{G}|}\sum_{\chi \in \hat{G}} \chi(\g) U^\dagger_v(\chi) U_e(\chi) U_{v'}(\chi) ,
\end{align}
for $\g\in G$ and $e$ directed from $v$ to $v'$.
The outcomes of this measurement are labelled by group elements $\g\in G$. 
The state resulting from this measurement can be written  
\begin{align}
    \ket{\Psi_{UG}}=\prod_{e\in R} \Pi_e(\g_e) \ket{+}_{ V}\ket{\Psi} ,
\end{align}
where $\ket{\Psi}$ is the original 1-form symmetric state which has support on the edges, $\ket{+}_{ V}=\ket{+}^{\otimes V}$ are the newly initialized states, and $\g_e\in G$ are the observed measurement outcomes on edges.  
The measured group variables $\g_e$ must form a 1-cochain denoted $\g^E\in C^1(R,G)$ such that $\g^E(e)=\g_e\in G$ on subregion $R$.  This is because the product of edge operators $U^\dagger_v(\chi) U_e(\chi) U_{v'}(\chi)$ around any face is an element of the 1-form symmetry group and hence has $\ket{\Psi}$ as an eigenstate with eigenvalue $+1$. 
Furthermore, since the region is simply connected, the 1-cochain $\g^E$ is in fact a 0-coboundary  $\g^E = \delta \g^V$ for some $\g^V\in C^0(R,G)$. 
Let $L(\g)$ denote left multiplication by $\g\in G$. One can equivalently choose the right multiplication in principle since $G$ is Abelian. 
Hence, there is some measurement outcome dependent byproduct operator $B(\g_v) = \prod_{v \in R} L_v(\g_v)$, such that 
\begin{align}
      \ket{\Psi_{UG}}= B(\g_v) \prod_{e\in R} \Pi_e(1) \ket{+}_{ V}\ket{\Psi} ,
\end{align}
which follows from the commutation relations
\begin{align}
    L_v(\g) \Pi_e(\h) = \Pi_e(\mathbf{gh}) L_v(\g) ,
    \\
     L_{v'}(\g) \Pi_e(\h) = \Pi_e(\overline{\g}\h) L_{v'}(\g) ,
\end{align}
for $e$ directed from $v$ to $v'$. 
After measuring and applying the correction operator as described below Eq.~\eqref{eq_character_p} we arrive at $B(\g_v)\ket{\Psi_{UG}}$ which is the result of ungauging the state $\ket{\Psi}$ in region $R$ with no residual domain walls due to the measurement outcomes. 
\begin{figure}[t]
    \centering
    \includegraphics[page=9]{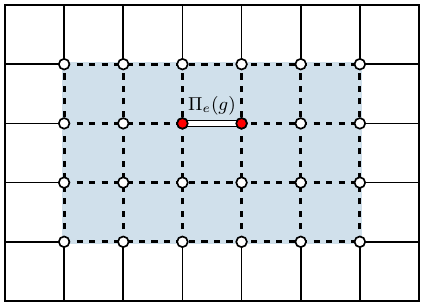}
  \caption{Gauging the 1-form symmetry. For simplicity, a square lattice is chosen for visualization. But the procedure works for a generic lattice. The subregion $R$ is colored in light blue-gray. The vertex ancillas are shown in small white circles in $R$, initialized in $\ket{+}$. We measure $\Pi_e(g)$ for every single edge in the region $R$ (the dashed-line sub-lattice). For example, one of the edge projector $\Pi_e(g)$ is shown as a double-line with red end points.}
    \label{figure_gauging_ungauging_a_region}
\end{figure}
%----
To regauge, we can simply measure the charge of each vertex degree of freedom and then remove them, as appropriate gauge field degrees of freedom on the edges are already present from the ungauging step. 
This is equivalent to measuring out each vertex degree of freedom in the character basis. 
To see the effects of this measurement we expand the ungauged state as follows
\begin{align}
    B(\g_v)\ket{\Psi_{UG}} = \frac{1}{|G|^E} \sum_{c \in C_1(R,\hat{G})} U(c) \ket{+}_{V}\ket{\Psi}
\end{align}
where $U(c)=U_E(c)U_V(\partial c)$
\begin{align}
    U_E(c) = \prod_e U_e(c_e), && U_V(\partial c) = \prod_v U_v( \{\partial c\}_v),
\end{align}
for $c$ a 1-chain on $R$. 
After measuring, the gauged state takes the form
\begin{align}
    \bra{\chi_v}_V \frac{1}{|G|^E} \sum_{c \in C_1(R,\hat{G})} U(c) \ket{+}_{V}\ket{\Psi}
\end{align}
where $\prod_v \chi_v = 1$ as the state being measured is symmetric under $\prod_{v} U_v(g)$.
The above state can be rewritten as: 
\begin{align}
\frac{1}{|G|^E} \sum_{\substack{c \in C_1(R,\hat{G}) \\ \text{s.t. } \partial c = \chi_v}} U_E(c) \ket{\Psi}. 
\end{align}
Since every chain $c \in C_1(R,\hat{G})$ that satisfies $\partial c = \chi_v$ can be rewritten as $c=c'+z$ for some cycle $z \in Z_1(R,\hat{G})$ and a fixed chain $c'$ satisfying $\partial c = \chi_v$, we can rewrite the state as
\begin{align}
    U_E(c') \frac{1}{|G|^E} \sum_{z \in Z_1(R,\hat{G})} U_E(c) \ket{\Psi}
    = U_E(c')\ket{\Psi} ,
\end{align}
as the sum is simply the projector onto the 1-form symmetric subspace in $R$ and we assumed the state $\ket{\Psi}$ was symmetric. 
Hence, up to a  on-site byproduct unitary operator, the ungauging and regauging procedure preserves any symmetric state. 

We remark that our procedure involves applying additional symmetric operators after ungauging. 
This does not alter the above result, as such operators can be commuted through the gauging map to equivalent gauged operators applied to the original state before ungauging~\cite{williamson2014matrix}. 

To demonstrate the ungauging and regauging procedure, we consider the toric code~\cite{qdouble}. In this case, the $\mathbb{Z}_2=\{0,1\}$ 1-form symmetry operators are given by products of $U_e(1)=Z$ operators on loops in a square lattice. 
The 1-form symmetric state $\ket{\Psi}$ is a ground state of the toric code. 
The left multiplication operator is $L(1)=X$. 
Ungauging implements a condensation of the vertex anyons. 
After ungauging, we find edge measurements that correspond to the coboundary of a vertex set. 
The byproduct operator is simply the product of $X$ on the vertices in this set. 
In the ungauged model, there is a global $\mathbb{Z}_2$ symmetry generated by applying $X$ to all vertices. 
After regauging this symmetry, the measurement outcomes correspond to vertex anyons, which can be paired up and annihilated via $Z$ operators on edge-paths connecting them. 

In summary, the general ungauging and regauging procedure described above can be applied to any 1-form symmetric state. This includes states that contain anyons with trivial 1-form charge, i.e.~those that braid trivially with the bosons that are condensed during the ungauging procedure. Ungauging such anyonic states suffices to implement the AFDLU procedure for general solvable anyon string operators as described in Section~\ref{solvable_anyon_string_op}.

Finally, we discuss the resources required to implement one round of the ungauging and regauging procedure. 
The ungauging procedure requires the addition of ancilla qudits on the vertices of the lattice, the measurement of three-body operators on edges and the surrounding vertices, and the application of a tensor product of unitary vertex operators. 
The circuit depth to implement the three-body measurements is determined by the coordination number of the lattice. 
The regauging procedure simply requires the measurement of edge degrees of freedom, followed by the application of a tensor product of unitary edge operators.

%-----
\subsubsection{Gauging anyon states}
\label{subsec:GaugingAnyonTubes}

We now illustrate the general gauging procedure by applying it to string-net states that contain anyonic excitations. This is a particular instance of the general gauging procedure, which was introduced above for the purpose of implementing solvable anyon string-operators.  We focus on a single term in the series of fusion categories $\mathcal{C}^{(i)}$. 
The  label $i$ is suppressed in this subsection.

Here we use extended string-net models that include dangling edges on each plaquette, which can be viewed as ancilla qudits, to allow violations of the fusion rules on vertices to be moved into plaquettes. 
Therefore, one can push the violations of the fusion rules % of string $\alpha$ 
at the end points of a string operator into corresponding plaquettes. 
We now generalize the controlled-plaquette operator $CB$ to include the quantum number $r$ of the tail of a string operator on a plaquette. 
This extends the gauging of the global $G$ symmetry to states that contain anyonic excitations. 
We define the tube operator~\cite{MR0996454,ocneanu1994chirality,Lan_2014,Bultinck2015,aasen2017fermion,williamson2017}
\begin{align}
    \mathcal{T}^{s_{\g}}_{pq_{\g}r}:=&\ \ 
\vcenter{\hbox{\includegraphics[page=15]{Figures.pdf}}}
 \ 
 = \ 
\vcenter{\hbox{\includegraphics[page=16]{Figures.pdf}}} ,
\label{eq_domain_wall_T_hexagon}
\end{align}
for $p,r\in \mathcal{C}_{\mathbf{1}}$ and $s_{\g},q_{\g}\in \C_{\g}$,
where the dashed line in the first equality represents periodic boundary conditions, so that this is an element of the defect tube algebra~\cite{williamson2017}. 
On the right of Eq.~\eqref{eq_domain_wall_T_hexagon}, the tube algebra element is depicted on a hexagon plaquette to demonstrate how the (defect) tube algebra acts on the string-net lattice.
Using the recipe in Ref.~\cite{williamson2017}, we define the following domain wall operators acting on anyons $a\in\mathcal{C}_{\mathbf{1}}^\text{rev}$, ($p,r\in \mathcal{C}_{\mathbf{1}}$)
\begin{align}
    \mathcal{B}^{\g}_a:=\sum_{pq_{\g}rs_{\g}} c^{s_{\g}}_{pq_{\g}r}(a)\mathcal{T}^{s_{\g}}_{pq_{\g}r} ,
    \label{eq_mathbfBg_def}
\end{align}
with coefficients $c^{s_{\g}}_{pq_{\g}r}(a)\in\mathbb{C}$, depending on $a$. 
The domain wall operators form a projective representation of the \textit{centralizer} subgroup $Z_a\leq G$ of group elements that do not permute the anyon type $a$
\begin{align}
    \B^{\h}_a \B^{\g}_a=  \eta_a(\h,\g) \B^{\h\g}_a,
    \label{B_op_projective_rep}
\end{align}
for $\h,\g\in\ Z_a$. 
Here $\B^{\mathbf{1}}_a$ corresponds to the irreducible central idempotent for the $a$ superselection sector~\cite{williamson2017}. 
With the generalized $(\mathcal{B}^{\g}_a)_p$ operators described above, one can gauge the generalized string-net including anyon and defect excitations on plaquettes via the following generalized KW duality 
\begin{align}
KW^{G}_{EP}:=& \bra{+}_P \prod_p (C\mathcal{B}_a)_p \ket{\{\chi_R \}}_P 
\label{eq_KW_EP_G} ,
\\
(C\mathcal{B}_a)_p\ket{\g}_p\ket{\Psi}:=&(\mathcal{B}_a^{\g})_{p}\ket{\g}_p\ket{\Psi} ,
\label{eq_CBg_psi}
\end{align}
where $\ket{\Psi}$ denotes the state of the $\mathcal{Z}(\C_{\mathbf{1}})$ topological order living on the edges $E$,  $\ket{\g}_p$ is the control ancilla at each plaquette $p$ supporting anyon type $a$ with $\g\in Z_a$, and $\ket{\{\chi_R\}}_P$ is defined below in Eq.~\eqref{eq_def_chi_R}. 
The label $a$ in $(C\mathcal{B}_a)_p$ is determined by the anyon type on plaquette $p$. 
Above, we cannot simply take the state corresponding to the trivial representation on plaquettes supporting anyons or defects as they may support projective representations where $\eta_a$ lies in a nontrivial second cohomology class from $H^2(Z_a,U(1))$. 
For this reason we use the plaquette state
\begin{align}
\ket{\{\chi_R \}}_P :=& \bigotimes_p \ket{\chi_{R_p}}_p ,
\label{eq_def_chi_R}\\
\ket{\chi_{R}}_p := & \frac{d_R}{|Z_a|^{\frac{1}{2}}} \sum_{\g \in Z_a} \chi_R^*(\g) \ket{\g} ,
\end{align}
where $R$ on a plaquette $p$ supporting anyon type $a$ is an irreducible projective representation of $Z_a$ with factor system $\eta_a$, and $d_R$ is the dimension of $R$. 
For plaquettes that support an anyon type that transforms under a linear representation of $Z_a$, such as the vacuum, we can take $R$ to be the trivial representation. 
In particular, when $G$ is a cyclic group, all anyons carry linear representations. 
To implement this gauging map, the plaquette degrees of freedom are again measured in the basis of linear irreducible representations of the abelian group $Z_a$. 
The plaquette tube projectors that result when gauging a plaquette supporting anyon $a$ with an outcome state $\bra{r}$, rather than $\bra{+}$, is
\begin{align}
     \Pi_{a,R\otimes r^*}=& \sum_{\h\in Z_a}\chi_R^*(\h)\chi_r(\h)\B_{a}^\h
     \\
     =&\sum_{\h\in Z_a}\chi_{R\otimes r^*}^*(\h)\B_{a}^\h ,
\end{align}
where $r$ is a linear rep of $Z_a$. 
The $\bra{r}$ outcome can be corrected to $\Pi_{a,R}$ via the application of a string operators to fuse a $\chi_r$ boson into the plaquette, as explained in Section~\ref{grading_and_ground_state_preparation}. 

Above we have chosen to introduce a $\mathbb{C}[Z_a]$ degree of freedom on a plaquette that supports anyon type $a$, rather than a full $\mathbb{C}[G]$ degree of freedom. 
When $Z_a$ is a proper subgroup, inequivalent anyons in the orbit of $a$ under the action of the global symmetry $G$ map to the same anyon sector after gauging. 
More specifically, the anyons described by projectors $\Pi_{^{\g}a,^{\g}R}$ become equivalent under the action of local operators in the gauged model $B_a^{\g}$ for $\g \in G / Z_a$ representative elements of the quotient group. 
Here, $^{\g}R$ denotes a projective irreducible representation of $Z_{{}^\g a}$ found by taking the conjugation action of $\g$ induced by $B_a^{\g}$ on the projective irreducible representation $R$ of $Z_{ a}$, see Ref.~\cite{williamson2017} for a detailed description. 

It is possible to modify the gauging map in the presence of anyons and defects to include a full $\mathbb{C}[G]$ degree of freedom on each plaquette. This is then decomposed into $\mathbb{C}[G/Z_a]\otimes \mathbb{C}[Z_a]$. 
After the initial gauging step in Eq.~\eqref{eq_KW_EP_G}, an additional step of applying the operator 
\begin{align}
\widetilde{KW}^{G}_{EP}:=& \bra{+}_P\qty(\prod_p (C\mathcal{B}_a)_p \ket{+}_P),
\end{align}
can be performed, where the plaquette states are associated with the quotient group variables $\mathbb{C}[G/Z_a]$ and hence plaquettes where $Z_a=G$ support no additional plaquette degree of freedom. 
After applying Eq.~\eqref{eq_KW_EP_G}, for a plaquette with anyon type $\Pi_{a,R}$, this results in an average over equivalent projectors $\Pi_{^{\g}a,^{\g}R}$
\begin{align}
     \sum_{\g \in G/Z_a} \Pi_{^{\g}a,^{\g}R}\ B_a^{\g} ,
\end{align}
where $a$ is permuted nontrivially by the global symmetry action being gauged. 
Correcting the observed linear rep outcomes of $G/Z_a$ from $\bra{r}$ to $\bra{+}$ is similar to the process described above. 

On each plaquette, the gauging procedure is implemented as follows. 
Before we act on the plaquette by $(\mathcal{B}_a^{\g})_p$, the tail of an excitation, labelled by a quantum number $r\in \mathcal{C}_{\mathbf{1}}$, is located at the centre of the punctured plaquette, depicted in dark gray in Eq.~\eqref{eq_tube_composition}. 
The tail $r$ can be viewed as an additional ancilla in plaquette $p$. 
Suppose the radius of the puncture is $r_1$.  We enlarge the puncture to radius $r_2>r_1$.  Let the tube $\Sigma$ that supports this defect tube algebra be defined by $\Sigma:=\{r_1\leq \norm{v}_2\leq r_2\}$.
We now apply the defect domain wall $B^{\g}_p$ to plaquette $p$ by composing the tube $\Sigma$, depicted in light gray in Eq.~\eqref{eq_tube_composition}, with the enlarged puncture, such that the outer boundary of the tube is sewed with the puncture, and the inner boundary  becomes the puncture of the updated hexagon plaquette.
\begin{align}
\vcenter{\hbox{\includegraphics[page=17]{Figures.pdf}}}
\ \  \cong \ \ 
\vcenter{\hbox{\includegraphics[page=18]{Figures.pdf}}}
 \ \ 
 \to 
 \ \ 
\vcenter{\hbox{\includegraphics[page=19]{Figures.pdf}}}
 \label{eq_tube_composition}
 \end{align}
Consequently, the tail is associated with a new quantum label $p$ on the other side of the $\g$-domain wall as shown in Eq.~\eqref{eq_domain_wall_T_hexagon} and~\eqref{eq_mathbfBg_def}. 
Using $KW_{EP}^G$ one can map the topological order of $\mathcal{Z}(\C_{\mathbf{1}})$ to $\mathcal{Z}(\C_G)$ including mapping the anyonic excitations. 
We illustrate this idea in detail for the $TY(\mathbb{Z}_3)$ example below. 

Using the above recipe, one can prepare a distant pair of potentially cyclic but solvable anyons on the lattice via AFDLU. 
%--------------

A generalization of the procedure above can be applied to gauge states containing twist defect excitations. 
Finding the irreducible central tube idempotents that describe anyons in a string-net model is closely related to constructing string operators that create those anyons, see Refs.~\cite{Christian2023,Green2023} for a review of this connection. 
In appendix~\ref{app_G_crossed} we describe a method to find irreducible central tube idempotents and string operators for anyons and defects in symmetry-enriched string-net models based on $G$-crossed modular input categories and their gauged variants.

\subsection{Example: Anyon preparation in $\mathcal{D}(S_3)$}
For a quantum double model $\mathcal{D}(G)$ of finite group $G$, the simple objects in $\Rep(\mathcal{D}(G))$ are labeled by a conjugacy class $C$ of $G$ and a representation $\rho$ of the centralizer group of $C$ \cite{bakalov_2001,Beigi_2011}. In the case of $S_3$, the anyons are shown in Table~\ref{table_of_S3_anyons}. This topological order can be obtained from the $\mathbb{Z}_3$ toric code by gauging the $\mathbb{Z}_2$ charge conjugation symmetry \cite{PhysRevB.100.115147,verresen2022}.

Here, and in the example below, we follow a simplified procedure to find AFDLU implementations of the string operators for non-Abelian solvabe anyons. Both of these examples are derived via gauging an Abelian symmetry on an Abelian topological order. Hence, their string operators only require a single round of ungauging and regauging to implement.

\begin{table}[t]
\centering
\begin{tabular}{|c|c|c|c|c|c|c|c|c|c|} \hline
Notations & $\wt{A}$ & $\wt{B}$ & $\wt{C}$ & $\wt{D}$ & $\wt{E}$ & $\wt{F}$ & $\wt{G}$ & $\wt{H}$  \\ \hline  
$(C,\rho)$ & e,1 & e,Sgn & e,Std & $\overline{s},1$ & $\overline{s}$,Sgn & $\overline{r}, 1$ & $\overline{r},[\omega]$ & $\overline{r},[\omega^*]$ \\ \hline
$Z(C)$ & $S_3$ & $S_3$ & $S_3$ & $\mathbb{Z}_2$ & $\mathbb{Z}_2$ & $\mathbb{Z}_3$ & $\mathbb{Z}_3$ & $\mathbb{Z}_3$ \\ \hline
dim & 1 & 1 & 2& $3$ & $3$ & $2$ & $2$ & $2$ \\ \hline
\end{tabular}
\caption{Table of anyons of $\mathcal{D}(S_3)$. The term Sgn refers to the sign representation in $S_3$ or $\mathbb{Z}_2$; the term Std refers to the 2-dimensional standard representation of $S_3$. For the group $\mathbb{Z}_3$ we use the generators to signify  the associated representations. The letter $\omega$ denotes the complex number $e^{2\pi i/3}$.}\label{table_of_S3_anyons} 
\end{table}

\begin{figure}[t]
    \centering
    \includegraphics[page=20]{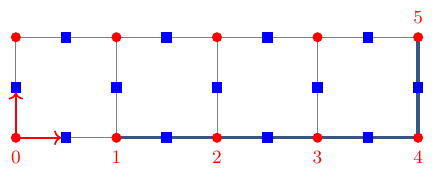}
    \caption{Part of the square lattice that supports the theory $\mathcal{D}(S_3)$. Here we only show four squares and part of the string $s$. The blue spin sitting on an edge $\langle ij\rangle$  is labelled by two vertices  $i$ and $j$ at its two ends.}\label{string_calculation_of_DS3}
\end{figure}

The anyons have an emergent $\mathbb{Z}_2$-grading inherited from the gauging process of $\mathbb{Z}_2=\{0,1\}$. 
The $\wt{D}$ and $\wt{E}$ anyons are assigned to the $1$-graded sector, while the remaining six anyons are assigned to the $0$-graded sector. 
To create a string of anyon type $\widetilde{C}\in \mathrm{Rep}(\mathcal{D}(S_3))$ within an area $A$ via AFDLU, we can first ungauge along a thickened ribbon to incude condensation. 
To achieve this we effectively apply  $\bra{0}_{\mathbb{Z}_2}$ projectors to the area $A$ via measuring $Z$ in $\mathbb{Z}_2\subset S_3$. 
This results in the topological order $\mathcal{D}(\mathbb{Z}_3)$ in the region $A$, see Figure~\ref{figure_gauging_ungauging_a_region}. 
Let $\ket{\Psi_0}$ denote the current state. The excitation in $\mathcal{D}(\mathbb{Z}_3)$ that relates to $\widetilde{C}\in \mathcal{D}(S_3)$ via gauging and ungauging is the electric excitation $e$ or its charge-conjugate $e^*$ (see, for example, Ref.~\cite{ren2023topologicalquantumcomputationassisted}). 
Either choice works equally well. 
Therefore, we create a string-$e$ operator supported on path $\gamma$ in  region $A$.
\begin{align}
    \ket{\Psi_1}:=W_s^{\mathbb{Z}_3} \ket{\Psi_0},\quad W_s^{\mathbb{Z}_3}=\prod_{\langle ij\rangle \in \gamma}\mathcal{Z}_{ij}.
\end{align}
Using the language in Refs.~\cite{verresen2022,Tantivasadakarn23Hierarchy}, we  then apply the gauging map $\mathcal{G}$ to the state on the region $A$
\begin{align}
  \ket{ \Psi_2}=&\mathcal{G}\ket{\Psi_1}:= KW_{EV}^{\mathbb{Z}_2} U_{EV}\ket{\Psi_1}\ket{+}_V^{\mathbb{Z}_2},\label{eq_gauging_of_S_3}\\
    U_{EV}=&\prod_v\prod_{v\to e} C_v^{\mathbb{Z}_2}\mathcal{C}_{e}^{\mathbb{Z}_3} .
    \label{eq_UEV_of_S_3}
\end{align}
Here, the product over $v\to e$ denotes all edges at vertex $v$ such that $v$ is the incoming vertex of the edge $e$ given the orientation of the lattice, see Figure~\ref{string_calculation_of_DS3}. The operator $C_v^{\mathbb{Z}_2}\mathcal{C}_e^{\mathbb{Z}_3}$ is the controlled-charge-conjugation gate with the control on the vertex qubit and the target on the edge qutrit. 
Finally, 
\begin{align}
    KW_{EV}^{\mathbb{Z}_2}=\bra{+}_V^{\mathbb{Z}_2} \prod_{\langle e,v\rangle } C_v^{\mathbb{Z}_2}X_e^{\mathbb{Z}_2}\ket{0}_E^{\mathbb{Z}_2} , \label{eq_KW_EV_S3}
\end{align} is the Kramers-Wannier map for the $\mathbb{Z}_2$ group with $C_v^{\mathbb{Z}_2}X_e^{\mathbb{Z}_2}$ the controlled-NOT for vertex $v$ and one of its connected edges $e$. 
To describe the resulting state  $\ket{\Psi_2}$ explicitly, it is equivalent to look at how the operator $W_s^{\mathbb{Z}_3}=\prod_{\langle ij\rangle \in s}\mathcal{Z}_{\langle ij\rangle}$ is mapped under the gauging procedure. 
Using $X,Z$ to denote $\mathbb{Z}_2$ operators and $\mathcal{X},\mathcal{Z},\mathcal{C}$ to denote $\mathbb{Z}_3$ operators, we have
\begin{align}
(C_v\mathcal{C}_e) \mathcal{Z} (C_v\mathcal{C}_e)=\mathcal{Z}^Z .
\end{align}
For notational simplicity we label the links $\langle i\, (i+1)\rangle$.  Without loss of generality, consider a string $s$ as shown in dark blue in Figure~\ref{string_calculation_of_DS3}. 

The action of the controlled charge conjugation $C_c\mathcal{C}_t$ is depicted in Figure~\ref{string_calculation_of_DS3} by an arrow from controlled qubit (a red dot)  to the target qutrit (a blue square). 
For each control qubit, it acts on two of the four neighboring target qutrits along the edges: the one above and the one to the right. Let $g(p,q):=r^p s^q$ be a group element in $S_3$, where $g$ can be viewed as a function written as a monomial of generators $r$ and $s$ of $S_3$.
Hence, given the $\mathbb{Z}_3$-string of the form $\mathcal{Z}_{12}\mathcal{Z}_{23}\mathcal{Z}_{34}\mathcal{Z}_{45}\cdots$, we have
\begin{align}
 & U_{EV}W_s^{\mathbb{Z}_3} U_{EV}^\dagger
  =\mathcal{Z}_{12}^{Z_1}\mathcal{Z}_{23}^{Z_2}\mathcal{Z}_{34}^{Z_3} \mathcal{Z}_{45}^{Z_4}\cdots\nnb\\
  =&(\mathcal{Z}_{12}\mathcal{Z}_{23}^{Z_1Z_2}\mathcal{Z}_{34}^{Z_1Z_3} \mathcal{Z}_{45}^{Z_1 Z_4}\cdots)^{Z_1}\nnb\\
   =&(\mathcal{Z}_{12}\mathcal{Z}_{23}^{Z_{12}}\mathcal{Z}_{34}^{Z_{12}Z_{23}} \mathcal{Z}_{45}^{Z_{12}Z_{23} Z_{34}}\cdots)^{Z_1}\nnb\\
  \to &   (\mathcal{Z}_{12}\mathcal{Z}_{23}^{Z_{12}}\mathcal{Z}_{34}^{Z_{12}Z_{23}} \mathcal{Z}_{45}^{Z_{12}Z_{23} Z_{34}}\cdots+h.c. ) \nonumber \\
  =& \sum_{p\in \mathbb{Z}_3}\sum_{q\in \mathbb{Z}_2}(\omega^{p} +\omega^{-p})P^{g(p,q)}_\gamma .
  \label{eq_UWZ3U_dagger}
%-------
\end{align}
Here, $P_\gamma^g=\delta_{\prod_{e\in \gamma} g_e,g}$ is a projection that demands the group multiplication of all the edge degrees of freedom along $\gamma$  equal to $g$,\footnote{There is a technical subtlety due to the fact that the orientation convention of the lattice and the direction of the string will revert some edge states $g_e$ to their group inverses $g_e^{-1}\in S_3$. However we have chosen a path of the string $\gamma$ so that we can avoid this subtlety.}
the third line  results from fact that the state after the action of $\prod_{\langle e,v\rangle}C_v^{\mathbb{Z}_2}X_e^{\mathbb{Z}_2}$ is stabilized by $Z_i^VZ_{ij}^EZ_j^V=1$ for each link $\langle ij\rangle$, and hence a pair of vertex Pauli $Z$ operators can be replaced by a product of edge Pauli $Z$ operators connecting them. The arrow in the fourth line is due to the measurement $\bra{+}_V$; namely the expression on the right-hand side is the resulting operator that commutes with the stabilizer $X_v$ for ancilla~$v$. 

Let $[F_\gamma]_{j_L,j_R}$ represent the string operator of anyon $\widetilde{C}\in \text{Rep}(\mathcal{D}(S_3))$, where $j_L,j_R$ label the internal degrees of freedom of the pair of anyons located at the left end $L=\partial_0 \gamma$ and the right end $R=\partial_1\gamma$, respectively. This object can be viewed as an operator-valued matrix. For example, its trace $\Tr[F_\gamma]$ is an operator acting on the Hilbert space of the lattice, rather than a $\mathbb{C}$-number.
In fact, the indices $j_L,j_R\in \{0,1\}$ correspond to the row and column indices of the 2-dimensional Standard matrix representation of $S_3$. We refer the interested reader to Appendix~B of Ref.~\cite{Bombin_2008} for more details about the $F$-string or ribbon operator. 
For our purposes, the right-hand side of Eq.~\eqref{eq_UWZ3U_dagger} is simply 
\begin{align}
  \sum_{p\in \mathbb{Z}_3}\sum_{q\in \mathbb{Z}_2}(\omega^{p} +\omega^{-p})P^{g(p,q)}_\gamma= \Tr[F_\gamma]+\text{ATr}[F_\gamma] ,
  \label{eq_tr_anti_trace}
\end{align}
where $\text{ATr}[M]=\sum_{j=1}^{n} M_{j,(n-j+1)}$ is the anti-trace of a matrix $M$.  This expression makes sense as we can label the independent internal spins of the two anyons at the ends of $\gamma$ via a pair of qubits $\{\ket{0}_{L/R},\ket{1}_{L/R}\}$.  The tensor product of plus states on both qubits qives
\begin{align}
    \ket{+}_L\ket{+}_R=(\ket{00}+\ket{11})_{LR}+(\ket{01}+\ket{10})_{LR}
\end{align}
which corresponds to the sum of trace and anti-trace, respectively. 
On the other hand, suppose instead of measuring $\bra{+}_V=\bigotimes_{v\in V}\bra{+}_v$ independently at every vertex $v\in V$ in Eq.~\eqref{eq_KW_EV_S3}, we measure $Z_{v_L}Z_{v_R}=+1$ and $X_{v_L}X_{v_R}=+1$ (note, here $Z_{v_L}\equiv Z_1$). Then there is a projection constraining $Z_{12}Z_{23}\cdots=+1$ that appears in the second last line of Eq.~\eqref{eq_UWZ3U_dagger}, which reduces the summation over $q\in \mathbb{Z}_2$ to only $q=0$. 
As the summation of the group element $g(p,q)=r^p s^q=r^p$ is restricted to not contain $s\in S_3$, we lose the anti-trace term from Eq.~\eqref{eq_tr_anti_trace}.  

We now discuss how this example connects to the procedure in Fig.~\ref{string_of_solvable_anyons}. Here, we have $a_{\g}=b_{\mathbf{1}}^{(i)}=\widetilde{C}$, while $c^{(i)}_{\h_i}$ and $c^{(i)}_{\h_{\leq i}}$ are $e$ or $e^*$ excitations of the $\mathbb{Z}_3$-toric code. The target anyon $\widetilde{C}$ lies in the identity graded sector of $\mathcal{Z}(S_3)$, and there is only one level of graded involved. Therefore, we do not need to create $O(n)$ pairs of $\widetilde{C}$ to implement the string operator. 
To summarize, we start from a pair of $e$ excitations supported on $\gamma$ in $\mathbb{Z}_3$-toric code, and end up with a pair of  $\widetilde{C}$ anyons at the two ends of $\gamma$ in a Bell pair $\frac{1}{\sqrt{2}}(\ket{00}+\ket{11})_{LR}$. 
This can be seen as a consequence of measuring $ZZ$ and $XX$ between the ends during the gauging protocol. 
We remark that this entanglement is susceptible to noise, corresponding to the fact that the internal degrees of freedom of anyons are local and not topological in nature. 
The anyon type of the string operator is a topologically robust quantity. 
Prescriptions for using such ribbon operators for topological quantum computation have been outline in Refs.~\cite{Cui_2015,nayak2008non, PhysRevX.12.021012}, for example. 
%---------

One can check that after the condensing and gauging procedure there exists no nontrivial domain wall on $\partial A$. 
For more background on this gauging procedure of $\mathcal{D}(S_3)$, we refer to Appendix~D of Ref.~\cite{verresen2022}.
%---------

\subsection{Example: Anyon preparation in $\mathcal{Z}(TY(\mathbb{Z}_3))$}
In this section, we discuss the ground state and string operators of the $\mathbb{Z}_3$ Tambara-Yamagami string-net.

\subsubsection{Explicit formula of the ground state of $\mathcal{Z}(TY(\mathbb{Z}_3))$}
We consider a honeycomb lattice on the spatial manifold under consideration. 
We first analyze the ground state of $\mathcal{D}(\mathbb{Z}_3)$ on an area $A$ of the spatial manifold, whose boundary is a contractible loop $\Gamma$ on the spatial lattice. 
The unnormalized ground state on the sublattice surrounded by $\Gamma$ is 
\begin{align}
\ket{\Omega}_{A}=\prod_p \sum_{a\in \mathbb{Z}_3} B_p^a\ket{0}_E=\sum_{\text{SN}}\ket{\text{SN}} ,
\end{align}
where each operator $B_p^a$ fuses $a\in \mathbb{Z}_3$ to the edges of the plaquette $p$. This state is the equal superposition of all diagrams or String-Nets  (SN) in $\Gamma$  satisfying $\mathbb{Z}_3$ fusion rules, as all $F$-symbols in $\mathbb{Z}_3$ are trivial. 

\begin{figure}[t]
    \centering
    \includegraphics[page=21]{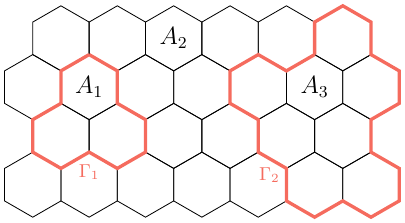}
    \caption{Illustration of the ground state wavefunction of $\mathcal{Z}(TY(\mathbb{Z}_N))$. Each red loop represents a $\sigma$-domain wall. }
    \label{illustration_of_the_ground_state_of_ZTYZ_N}
\end{figure}

\begin{figure}[t]
     \centering
    \includegraphics[page=22]{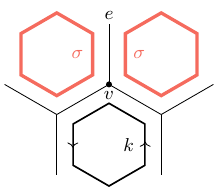}
     \caption{A tensor network point of view on three plaquette operators acting on $\ket{0}_E$ on the honeycomb lattice. 
     A pair of $\sigma$-loops (red) are glued along their shared edge $e$. 
     The vertex $v$ is one of the two vertices of edge $e$, referred to as the gluing-point in the main text.}\label{fig_two_plaquette_sigma_loop}
\end{figure}

We now discuss the Tambara-Yamagami category~\cite{gelaki2009centers,TAMBARA1998692} $TY(\mathbb{Z}_N,\chi,\pm )$ for the cyclic group $\mathbb{Z}_N$. 
The label $\chi$ is a bicharacter of the group, and $\pm$ indicates the choice of a sign. In this work, we restrict our attention to $N=3$, $\chi(a,b)=\omega^{ab}$ with $\omega=e^{2\pi i/3}$, and the plus sign $+$. 
There are four simple objects in this category, including three group elements $a\in \mathbb{Z}_3$ and a non-invertible object $\sigma$. 
The fusion rules include
\begin{align}
    &a\otimes b=a+b\mod 3,\quad a\otimes \sigma=\sigma \otimes a=\sigma,\\
    &\sigma\otimes \sigma=0\oplus 1\oplus 2.
\end{align}
The nontrivial $F$-symbols occur when there are four $\sigma$ strings among the indices of the $F$-symbol:
\begin{align}
F^{a\sigma b}_{\sigma\sigma\sigma}=F^{\sigma b \sigma}_{a\sigma\sigma}=\omega^{ab},\quad F^{\sigma\sigma\sigma}_{\sigma ab}=\omega^{-ab},
\end{align}
otherwise the $F$-symbols are either 0 or 1. 
The Tambara-Yamagami category supports a $\mathbb{Z}_2$ grading, which is generated by its unique non-invertible simple object $\sigma$.  Therefore, the picture of the ground state wavefunction is similar to the other $\mathbb{Z}_2$-graded examples we have considered, namely, the toric code and the Ising topological order. 
By definition, the ground state of the string-net model of $TY(\mathbb{Z}_3)$ is given up to normalization by 
\begin{align}
   \ket{\Omega}_{TY(\mathbb{Z}_3)}=\prod_p \sum_{s\in TY}d_s B_p^s\ket{0}_E ,
\end{align}
where the dimension is $d_a=1$ for $a\in \mathbb{Z}_3$ and $d_\sigma=\sqrt{|\mathbb{Z}_3|}=\sqrt{3}$. 
In contrast to the ground state of $\mathcal{D}(\mathbb{Z}_3)$, the string-net coefficients in this case are nontrivial.  
Each string-net configuration in $TY(\mathbb{Z}_3)$ can be viewed as consistent $\mathbb{Z}_3$ string-nets partitioned by $\sigma$-domain walls, with the coefficient depending only on the domain wall configurations. 
Consider a TY SN configuration with $\sigma$-loops supported on loops $\Gamma=\{\Gamma_i\}$ with each loop labelled by $i$.
Let $\{A_j\}$ be the areas on the spatial manifold, partitioned by $\Gamma$ as shown in Figure~\ref{illustration_of_the_ground_state_of_ZTYZ_N}. Then the ground state wavefunction can be formulated as
\begin{align}
    \ket{\Omega}_{TY(\mathbb{Z}_3)}=\sum_{\Gamma }\varphi(\Gamma) \bigotimes_i\ket{\Gamma_i}\bigotimes_j\ket{\Omega}_{A_j} ,
\end{align}
where $\ket{\Gamma_i}$ is a product of states $\ket{\sigma}$ supported on $\Gamma_i$, and 
the coefficient $\varphi(\Gamma)=\prod_i \varphi(\Gamma_i)$ is the product of factors, $\varphi(\Gamma_i)$,  over $i$. 
To evaluate $\varphi(\Gamma_i)$, we note that each $\sigma$-loop can be constructed by fusing multiple smaller $\sigma$-loops, each supported on a single plaquette.  For example, consider a $\sigma$-loop that spans two plaquettes as shown in Figure~\ref{fig_two_plaquette_sigma_loop}.

Suppose the two $\sigma$ labels fuse to $b\in\mathbb{Z}_3$ on edge $e$. 
We assign the following gluing-point factor $g_v$ to the vertex $v$ along $\Gamma_i$:
\begin{align}
   g_v=\omega^{-kb} ,\quad \omega=e^{2\pi i/3} ,\label{gluing_factor}
\end{align}
which is derived from the following F-move:
\begin{align}
\vcenter{\hbox{\includegraphics[page=23]{Figures.pdf}}}
    =F^{k\sigma (-b)}_{\sigma;\sigma\sigma}\sqrt{\frac{d_\sigma d_k}{d_\sigma}}
\vcenter{\hbox{\includegraphics[page=24]{Figures.pdf}}}
    .
    \label{eq_fusing_two_sigma_plaquettes}
\end{align}
The LHS of Eq.~\eqref{eq_fusing_two_sigma_plaquettes} corresponds to the layout of Fig.~\ref{fig_two_plaquette_sigma_loop} exactly, where two sigma-loops and and $k$-loop is fused to the three plaquette and $b\in \mathbb{Z}_3$ is the fusion product of $\sigma\otimes \sigma$ supported on the shared edge $e$ in Fig.~\ref{fig_two_plaquette_sigma_loop}.  We recommend the reader the original reference \cite{Levin_2005} by Levin and Wen for more details.
%------------
For an orientation that is different from that shown above, we apply charge conjugation to $k$ or $b$. 
We then compute $\varphi(\Gamma_i)$ by taking the product of all the gluing-point factors on the boundary $\Gamma_i$
\begin{align}
    \varphi(\Gamma_i)=\prod_{v\in \Gamma_i}g_v.
\end{align}
To prepare the ground state using a finite depth circuit, we adopt the same procedure as in Sec.~\ref{sect_the_second_gauging}. Namely,
\begin{align}
    \ket{\Omega}_{TY(\mathbb{Z}_3)}=\bra{+}_P \prod_{p\in P}CB_p\ket{\Omega_{\mathbb{Z}_3}}\ket{+}_P.\label{gauging_Z_3_to_TY}
\end{align}
%-------
\subsubsection{Anyon preparation in $\mathcal{Z}(TY(\mathbb{Z}_3))$}
In this section we deal in detail with plaquette and string operators of the string-net model. 
There are fifteen anyons in the center of $TY(\mathbb{Z}_3)$ (see Section~VII.E in Ref.~\cite{PhysRevB.103.195155}). 
We focus solely on the anyon $\Phi=\mathbf{1}\boxtimes \phi\in \mathcal{C}_1\boxtimes \mathcal{C}^\mathrm{rev}_G$, which is $\alpha=9$ in Ref.~\cite{PhysRevB.103.195155}. 
This is because $\Phi$ has a cyclic fusion rule $\Phi^2=1+Z+\Phi$, $Z=\mathbf{1}\boxtimes z$, and hence $\Phi$ is one of the most challenging to prepare. 

The relevant half-braiding data can be found in Refs.~\cite{PhysRevB.100.115147} and \cite{PhysRevB.103.195155}
\begin{align}
    \Omega_\Phi^{1,112}=&\Omega_\Phi^{2,221}=\omega^{*}\quad 
    \Omega^{1,220}_\Phi=\Omega_\Phi^{2,110}=\omega \label{Z_3_part_of_the_string_data}\\
\Omega_\Phi^{\sigma,12\sigma}=&\omega^{2-n},
\quad \Omega_\Phi^{\sigma,21\sigma}= \omega^{2+n} , \label{toll_factor}
\end{align}
%--------
where all other entries of $\Omega$ are zero. 
The parameter $n$ is a gauge choice, and therefore, we may set $n=0$.
We now describe the wavefunction of a state containing a pair of excitations. 
Such a state can be created by a string operator applied to the ground state. 
Let $W_\gamma$  be the string operator on an oriented path $\gamma$ from $\gamma_0$ to $\gamma_1$. Such an operator can be decomposed into $W_\gamma^r$ with $r$ labelling the simple object of a leg at the initial plaquette $\gamma_0$. We show below that there is no need for a second label for the leg at plaquette $\gamma_1$, since it can be derived from $r$. 
Then we have 
\begin{align}
    W_\gamma^\Phi\ket{\Omega}= \sum_{r\in TY(\mathbb{Z}_3)}W_\gamma^r\ket{\Omega}.
\end{align}
The half-braiding data $\Omega_\Phi$ of $\Phi$ restricts $r$ to  be either 1 or 2 in $\mathbb{Z}_3$.  To evaluate the action of $W_\gamma^r$  on $\ket{\Omega}$, it is sufficient to work out the action on a specific configuration 
\begin{align}
 \ket{\psi}:=\bigotimes_i\ket{\Gamma_i} \bigotimes_j\ket{\Omega}_{A_j} ,
\end{align} 
with $\{\Gamma_i\}$ and $\{\ket{\Omega}_{A_j}\}$ defined above.
Let 
\begin{align}
\{ \gamma_0,\gamma_1,\gamma_2,\cdots ,\gamma_N \},
\end{align} 
be a partition of the path $\gamma$ that is segmented by $\sigma$-loops. Assign to each $\gamma_k$ the string label $(-1)^k r$. 
Each time the path $\gamma$ passes through a $\sigma$-domain-wall, we negate the value of $r$ in $\mathbb{Z}_3$. For simplicity, label the corresponding area containing $\gamma_k$ as $A_k$, ignoring areas that do not contain any part of $\gamma$. We denote the crossings in Eq.~\eqref{eq_half_braiding1} and Eq.~\eqref{eq_half_braiding2} as type 1 and 2, respectively.
Let $N_1$ ($N_{2}$) denote the number of the type 1 (2) crossings of the string $\alpha$ with $\sigma$-domain walls.
Then we have
\begin{align}
 W_\gamma^r    \ket{\psi}=\omega^{2(N_1-N_2)}\qty(\bigotimes_i\ket{\Gamma_i}) \qty(\bigotimes_k W_{\gamma_k,\mathbb{Z}_3}^{(-1)^k r}\ket{\Omega}_{A_k}) ,
 \label{r_string_op_on_TY_background}
\end{align}
where $W^s_{\gamma_k,\mathbb{Z}_3}$ is the $s$-string operator on the ground state of $\mathbb{Z}_3$ toric code, described explicitly in Eq.~\eqref{Z_3_part_of_the_string_data}. 
Additionally, the half-braiding $\Omega_\Phi^{a,rr(r+a)}=\omega^{-ra}$ characterizes the string operator $W_{\gamma}^{em^*}$ of $em^*$ for $r=1$ and the string operator $W^{e^*m}_\gamma$ of $e^*m$ for $r=2$.
This action factorizes over areas that are partitioned by $\sigma$-lines and each time $\gamma$ passes a $\sigma$-line, there is a flip $r\to -r$ and a factor $\omega^{\pm2}$ due to Eq.~\eqref{toll_factor}. 
Eq.~\eqref{embar_string_on_Z_3} illustrates the  action of $W^r_{\gamma,\mathbb{Z}_3}$ directly on a the background configuration that is locally a ground state configuration of the $\mathbb{Z}_3$ string-net. 
The fusion with the $r$-string originates from the magnetic sector of $\mathbb{Z}_3$, while the phase $\omega^{r(m-i)}$ arises from the electric sector. 
The phase depends only on the value of the loop at the ends of the string $\alpha$, as the phases along the bulk of the string cancel out.
\begin{widetext}
\begin{align}
 W_s^r\ket{\Omega}_{\mathbb{Z}_3}=
\vcenter{\hbox{\includegraphics[page=25]{Figures.pdf}}}
=
\omega^{r(m-i)}
\vcenter{\hbox{\includegraphics[page=26]{Figures.pdf}}}
\label{embar_string_on_Z_3}
\end{align}
\end{widetext}
%-------
We now check that, following the protocol in Eq.~\eqref{gauging_Z_3_to_TY} the state  $(W_\gamma^{em^*}+W_\gamma^{e^*m})\ket{\Omega}_{\mathbb{Z}_3}$ is gauged and mapped to  $W_\gamma^\Phi$.

Solving Eq.~\eqref{eq_mathbfBg_def} and \eqref{B_op_projective_rep}, one arrives at 
\begin{align}
    c^s_{pqr}=
\begin{cases}
\frac{1}{3}\delta_{p,r}\delta_{q,r+s}\omega^{-rs},  &s\in \mathbb{Z}_3 ,\\
\frac{1}{\sqrt{3}}\delta_{p,-r} \delta_{q,\sigma} \omega^{2}, &s=\sigma .
\end{cases}
\label{eq_c_spqr}
\end{align}
where $\omega=e^{2\pi i/3}$. 
%-----------
Using the above data with the procedure introduced in Eqs.~\eqref{eq_mathbfBg_def} to  \eqref{eq_CBg_psi}, one can correctly map the $\mathbb{Z}_3$ topological order to $TY(\mathbb{Z}_3)$. This maps a $em^*+e^*m$ anyon string to a $\Phi$ anyon string in constant depth, where $\Phi$ corresponds to $a_{\g}$ and $b_{\mathbf{1}}^{(i)}$ in Fig.~\ref{string_of_solvable_anyons}. Upon measuring the resulting anyon charges, $em^*$ and $e^*m$ correspond to the $c^{(i)}_{\h}$ in Fig.~\ref{string_of_solvable_anyons}. Similar to the 
$\mathcal{D}(S_3)$ example above, the anyon $\Phi$ lies in the identity graded sector of the theory and only a single gauging step is required. Hence, it is not necessary to create $O(n)$ pairs of $\Phi$ anyons to implement the string operator considered here. 
This can be verified on the lattice through explicit calculation, by using the $\Omega_\Phi$ data given in Eq.~\eqref{Z_3_part_of_the_string_data}-\eqref{toll_factor}.
From Eq.~\eqref{eq_P_1} and Eq.~\eqref{eq_P_Phi}, one also finds that the controlled plaquette operator, followed by the projector $\bra{+}_P$, implements $P_{\mathbf{1}}+P_\Phi$.

%---------------------
Similar to the doubled Ising anyon example, 
the controlled plaquette operator
\begin{align}
    (CB_p)^\dagger (CB_p)=id\otimes \frac{1}{3}(B^0_p+B^1_p+B^2_p) ,
\end{align}
projects the state onto the subspace defined by $P_p^+:=\frac{1}{3}(B^0_p+B^1_p+B^2_p)$. So, formally, one needs an additional unitary to extend the operator to $\overline{CB_p}$ in the orthogonal subspace $(\mathds{1}-P^+_p)$, see the discussion in Appendix~\ref{app:CBCircuit}. 

In summary, an advantage of the gauging procedure, beyond the preparation of the topological order, is that it can generate $\Phi$. 
Namely, a naive fusion process to move $\Phi$ is probabilistic, because the fusion of a pair of $\Phi$ may result again in $\Phi$. However, using gauging, one can first prepare an $em^*+e^*m$ string and then gauge it to find a string of $\Phi$ via AFDLU. 
%---------

\section{Discussion}
\label{sec:Discussion}
In this work, we have constructed AFDLUs to prepare topological ground states for all solvable anyon theories that admit gapped boundaries.
Solvable anyon theories are a vast generalization of finite solvable groups and include anyons with non-integer quantum dimensions such as doubled Ising or $SU(2)_4$ anyons. 
Our approach was based on implementing sequential gauging of finite Abelian symmetries, following Refs.~\cite{TantivasadakarnmeasureSPT,verresen2022,Tantivasadakarn23Hierarchy}, where the discussion was limited to twisted solvable groups. 
Furthermore, we have extended this approach to construct AFDLUs that implement string operators of arbitrary length for solvable anyons and their symmetry twist defects. 
Our approach for implementing these string operators takes advantage of the structure of solvable anyon fusion and AFDLU implementations of gauging and ungauging 1-form symmetries to implement and reverse anyon condensation, respectively. 
The AFDLU string operators we find differ from those of Ref.~\onlinecite{bravyi2022adaptive} which were limited to quantum doubles of solvable groups and did not make use of gauging. 

In Ref.~\onlinecite{Tantivasadakarn23Hierarchy}, it was conjectured that the complete classification of the trivial topological phase under adaptive finite depth quasi-local unitaries (``measurement-equivalent phases") is given by all solvable anyon theories. In this work, we provide an explicit protocol to prepare all solvable string-net ground states, which correspond to solvable anyon theories that admit a gapped boundary.
\textit{We conjecture that solvable string-nets provide a complete classification of all 2D topological phases that can be created exactly via (strictly local) AFDLU.} 
This conjecture follows iff 2D AFDLUs are restricted to implement gauging and ungauging of global Abelian symmetries on topological phases. 
If our conjectured classification is not correct, there must exist AFDLUs that implement transformations beyond Abelian gauging and hence can be used to construct more general topological phases. 

Our results raise a number of questions for future work. 
First, can one find a proof of the above conjecture, or potentially find an adaptive circuit depth lower bound for a particular family of nonsolvable anyon states. 
An interesting unsolved challenge is to establish a firm connection between braiding universality and adaptive circuit depth lower bounds for anyon theories. 
Alternatively, can one find an AFDLU that prepares any example of a nonsolvable anyon theory such as doubled Fibonacci anyons or the double of a nonsolvable group, e.g. $A_5$. 
We remark that if one applies the ungauging and regauging AFDLUs we have developed to nonsolvable anyon theories, it is possible to reach different, AFDLU-equivalent, nonsolvable theories, but not the trivial theory. 
Second, how does the proposed classification change for approximate preparation via AFDLU. 
Third, can our protocols for AFDLU anyon preparation be made fault tolerant by incurring a larger spacetime overhead. 
This is an important question for any potential scalable practical applications of our results. 
In this direction, it would be interesting to extend the result of Ref.~\cite{Dauphinais2016} from nilpotent to solvable anyons. 
Alternatively, it would be interesting to further develop autonomous schemes for stabilizing non-Abelian anyons following Ref.~\cite{Chirame2024}. 
Fourth, what generalizations of our results to higher dimensions, including fracton phases, are possible. 
Can such phases be classified via topological defect networks~\cite{Aasen20} built from solvable constituents. 
Are all fracton phases AFDLU equivalent to layers of 2+1D and 3+1D conventional topological order~\cite{Vijay2016,Williamson2016Fractal,shirley2018FoliatedFracton,Williamson2020Designer}. 
Fifth, what is the mathematical classification of anyon theories up to gauging and ungauging Abelian symmetries. 
For Abelian anyon theories, this classification should match the Witt group~\cite{Davydov2013a,Davydov2013b}. 
For more general non-Abelian anyon theories, we expect the classification to be a monoid that is a generalization of the categorical Witt group~\cite{Davydov2013a,Davydov2013b} which only allows for gauging and ungauging of Abelian symmetries, rather than general finite group symmetries. 
Finally, it would be interesting to determine the minimum number of measurement rounds required to prepare a solvable string-net ground state. For example, in Refs.~\cite{Tantivasadakarn23shortestroute,Tantivasadakarn23Hierarchy} it was noticed that the $D_4$ quantum double can be prepared by one round of measurement by viewing it as a twisted $\mathbb Z_2^3$ quantum double. This question can now be precisely formulated using the language of fusion categories as follows: given a cyclically nilpotent fusion category, what is the smallest nilpotency class among all its Morita equivalent fusion categories that are also cyclically nilpotent?

Finally, we comment on the feasibility of implementing our protocols on current quantum devices. Our work directly generalizes the protocols in Refs.~\cite{Tantivasadakarn23shortestroute,Tantivasadakarn23Hierarchy}, which have been used to explicitly prepare the $D_4$ quantum double in a trapped ion quantum processor in Ref.~\cite{iqbal2024non}. Our generalization goes beyond twisted quantum double models, the simplest example being the doubled Ising model. It would thus be interesting to apply our protocols to realize the double Ising topological order. Given the successful preparation of the $D_4$ quantum double in existing quantum hardware, such an experiment appears to be tractable in the near term. Going beyond the doubled Ising model, a similar discussion applies to the $\mathbb{Z}_3$ Tambara-Yamagami model, see Appendix~\ref{app:CBCircuit} for further details about the gates required to implement the aforementioned models. 

%---------
\acknowledgements
We are grateful to the authors of Ref.~\onlinecite{lyons-2024} for informing us of their upcoming work, which also discusses efficiently creating anyons using adaptive circuits.
We thank Anasuya Lyons, Corey Jones, and Ruben Verresen for useful discussions.
This work was initiated at Aspen Center for Physics, which is supported by National Science Foundation grant PHY-1607611.
NT is supported by the Walter Burke Institute
for Theoretical Physics at Caltech.
Parts of this work were completed while DJW was visiting the Simons Institute for the Theory of Computing and the Kavli Institute for Theoretical Physics.
DJW was supported in part by the Australian Research Council Discovery Early Career Research Award (DE220100625). 
This material is partially based upon work supported by the U.S.
Department of Energy, Office of Science, National Quantum Information
Science Research Centers, Quantum Systems Accelerator, under Grant
number DOE DE-SC0012704.
This research was supported in part by the National Science Foundation under Grant No. NSF PHY-1748958. 
%--------
\bibliography{main.bib}
%-------
\appendix
%----------
\section{Review of string-net models}
\label{app_StringNet}

\subsection{String-net models and string operators}
In this section we briefly review the string-net models introduced by Levin and Wen~\cite{Levin_2005}, see also Ref.~\cite{PhysRevB.103.195155}. 
The string-net model is based on an input fusion category $\mathcal{C}$. 
Here, we describe string-net models with no fusion multiplicity, for simplicity of presentation. 
Our results in the main text apply to string-net models with fusion multiplicity and directed edges. 
The string-net model is defined on a trivalent lattice,  $\Lambda$, which is conventionally taken to be a honeycomb lattice.
The total Hilbert space is a tensor product of local Hilbert spaces  $\mathcal{H}=\bigotimes_{e\in \Lambda} \mathcal{H}_e$ with each $\mathcal{H}_e$ associated to an edge on the lattice. Each local Hilbert space is spanned by a basis of simple objects, or string types, in $\mathcal{C}$, 
\begin{align}
\mathcal{H}_e=\mathbb{C}[\operatorname{Simples}(\mathcal{C})], 
\end{align}
below we compress the above notation to $\mathbb{C}[\mathcal{C}]$. 
The lattice Hamiltonian is  
\begin{align}
   H^{\mathcal{C}}_\mathrm{SN}=-\sum_v A_v-\sum_p B_p,
   \label{eq_sn_hamiltonian}
\end{align}
where $v$ and $p$ label vertices and  plaquettes of $\Lambda$, respectively. 
We use the notation $\ket{abc}_v\in \bigotimes_{e\ni v}\mathcal{H}$ to denote the basis of edge states adjacent to vertex $v$. 
The vertex term $A_v$ is defined by acting on basis vectors as a projector 
\begin{align}
    A_v\ket{abc}_v=\delta_{ab}^c\ket{abc}_v,
\end{align} 
where the Kronecker-delta  $\delta_{ab}^c$ is nonzero only if $c$ is a possible fusion result of $a\otimes b$. 
This definition assumes the $a,b$ edges are oriented towards $v$, and $c$ is oriented away, edge orientations can be reversed up to swapping objects with their inverses. 
The plaquette term is
\begin{align}
    B_p=\sum_{a\in \mathcal{C}} \frac{d_a}{\mathcal{D}^2} B^a_p,
\end{align}
where $d_a$ is the quantum dimension of object $a$, ${\mathcal{D}^2=\sum_a d_a^2}$ is the total quantum dimension squared, and each plaquette string operator, $B_p^a$, fuses the string of type $a$ into plaquette $p$, using the $F$-symbol data of the fusion category $\mathcal{C}$.

The anyon theory that describes the emergent superselection sectors of the string-net model $\mathrm{SN}(\mathcal{C})$ is the modular tensor category $\mathcal{Z}(\mathcal{C})$, the Drinfeld center of $\mathcal{C}$. 
For example, the anyons of $\mathrm{SN}(TY(\mathbb{Z}_3))$, containing four simple objects, are given by $\mathcal{Z}(TY(\mathbb{Z}_3))$, consisting of fifteen different anyons.
For input categories $\mathcal{C}$ that admit a modular braiding, the center is simply the double of the input category $\mathcal{Z}(\mathcal{C})=\mathcal{C}\boxtimes \mathcal{C}^{\mathrm{rev}}$, where the simple objects are labeled by objects in $\mathcal{C}$ equipped with the modular braiding and its time reversed counterpart, respectively. 
Examples include $\mathrm{Vec}_{\mathbb{Z}_3}$, the double Ising model, and the double Fibonacci model~\cite{KOENIG20102707}.

We now describe the lattice string operators for the emergent anyons. 
For each simple object $\alpha\in \mathcal{Z}(\mathcal{C})$, there is a set of data $\Omega_\alpha$, known as a \emph{half-braiding}, which can be used to construct the relevant string operator. 
We use the notation $W^\alpha_\gamma$ for the string operator that creates the pair of anyons, $\alpha$ and its antiparticle $\overline{\alpha}$, at the two ends of the edge-path $\gamma$, respectively. 
A segment of the string operator is represented by a double line in Eq.~\eqref{eq_appendix_half_braiding1} and~\eqref{eq_appendix_half_braiding2}. 
The half-braiding data specifies how the $\alpha$-string is fused into the SN lattice via the following equations~\cite{PhysRevB.103.195155} 
\begin{align}
    \vcenter{\hbox{\includegraphics[page=36]{Figures.pdf}}}
=&\sum_{b,s,r}\Omega_\alpha^{a,rsb}
\ \ \sqrt{\frac{d_b}{d_a\sqrt{d_rd_s}}}\ \ 
\vcenter{\hbox{\includegraphics[page=12]{Figures.pdf}}}
\label{eq_appendix_half_braiding1}
\\
%-----------------
\vcenter{\hbox{\includegraphics[page=37]{Figures.pdf}}}
=&\sum_{b,s,r}(\Omega_\alpha^{a,srb})^*
\sqrt{\frac{d_b}{d_a\sqrt{d_rd_s}}}\ 
\vcenter{\hbox{\includegraphics[page=14]{Figures.pdf}}}
\, .
\label{eq_appendix_half_braiding2}
\end{align}
The half-braiding data $\{\Omega_\alpha^{a,rsb}\}$ defines how the crossing of the $\alpha$-string with a lattice edge, depicted as a circle above, can be implemented via an operator. 
In appendix~\ref{app_G_crossed} we discuss solutions to the half-braiding equations derived from modular braidings on an input category, see also Ref.~\cite{KOENIG20102707}. 

\subsection{Symmetry-enriched string-net models}

We now review the symmetry-enriched string-net models~\cite{PhysRevB.94.235136,cheng2016exactly}. 
These models are based on $G$-graded unitary fusion categories and describe all non-anomalous symmetry-enriched gapped topological phases of matter that admit gapped boundaries. 
A $G$-graded fusion category is a fusion category with a decomposition of simple objects into $G$-sectors
\begin{align}
    \mathcal{C}_G = \bigoplus_{\g \in G} \mathcal{C}_\g ,
\end{align}
where $G$ is a finite group. 
We use the notation $a_\g$ to denote $a\in \mathcal{C}_\g$. 
The fusion rules respect the grading in the following sense, if $c$ appears in the fusion product of $a_\g \otimes b_\h$, then $c\in \mathcal{C}_{\mathbf{gh}}$ and we write $c_{\mathbf{gh}}$. 

The plaquette operators, $B_p^a$, corresponding to the category $\mathcal{C}_G$ can be organized into sums that form a representation of $G$
\begin{align}
    B_p^{\g}:=\frac{1}{\mathcal{D}_i^2} \sum_{a\in \C_{\g}}\ d_aB_p^{a},\quad \g\in G,
    \label{eq_B_p_grading_sectors_appendix}
\end{align}
which satisfy $B_p^{\g}B_p^{\mathbf{h}}=B_p^{\mathbf{gh}}$ and $B_p^{\g^{-1}}=(B_p^{\g})^\dagger$.

The Hilbert space of the symmetry-enriched string-net is obtained by taking the tensor product of the Hilbert space for the string-net model based on $\mathcal{C}_G$ with additional $\mathbb{C}[G]$ degrees of freedom on every plaquette $p$ in the lattice. 
The Hamiltonian is 
\begin{align}
    H_{G\mathrm{SN}}^{\mathcal{C}_G} = -\sum_v A_v - \sum_p \frac{1}{|G|} \sum_{\g\in G} R_p(\g) B_p^{\g} - \sum_e C_e ,
\end{align}
where the sums are over vertices, plaquettes, and edges of a trivalent latice. 
The vertex terms are identical to those for a conventional string-net model based on the fusion category $\mathcal{C}_G$, and energetically enforce the fusion rules. 
The plaquette terms are given by products of right group multiplication operators $R(\g)\ket{\h}=\ket{\mathbf{hg}^{-1}}$ on the $\mathbb{C}[G]$ plaquette degree of freedom with the plaquette operators defined in Eq.~\eqref{eq_B_p_grading_sectors_appendix}. 
These terms fluctuate domains of $G$ variables on plaquettes, along with strings from the matching $\mathcal{C}_{\g}$ sector on their boundaries. 
The edge terms $C_e$ act diagonally in a basis of group variables on plaquettes and simple objects on the edges
\begin{align}
C_e\ \begin{tikzpicture}[baseline={(current bounding box.center)}]
\def\r{0.7};
\def\acolor{Teal};
\foreach \x in {0,1} {
      \node[regular polygon, regular polygon sides=6, draw, minimum size=2*\r cm, rotate=30] at (\x*1.732*\r, 0*\r) {\ };}
\node at (0,0.4*\r) {{\color{\acolor}$\g$}};
\filldraw[draw=\acolor,fill=\acolor] (0,0) circle[radius=0.07*\r];
\node at (0.5*1.732*\r,-0.9*\r) {{\color{\acolor}$a_{\h}$}};
\node at (1.732*\r,0.4*\r) {{\color{\acolor}$\k$}};
\filldraw[draw=\acolor,fill=\acolor] (1.732*\r,0) circle[radius=0.07*\r];
\draw[Teal,line width=2 pt] (0.5*1.732*\r,0.5*\r) -- (0.5*1.732*\r,-0.5*\r);
\end{tikzpicture}
=
\delta_{\g \h \k^{-1}}
\ 
\begin{tikzpicture}[baseline={(current bounding box.center)}]
\def\r{0.7};
\def\acolor{Teal};
\foreach \x in {0,1} {
      \node[regular polygon, regular polygon sides=6, draw, minimum size=2*\r cm, rotate=30] at (\x*1.732*\r, 0*\r) {\ };}
\node at (0,0.4*\r) {{\color{\acolor}$\g$}};
\filldraw[draw=\acolor,fill=\acolor] (0,0) circle[radius=0.07*\r];
\node at (0.5*1.732*\r,-0.9*\r) {{\color{\acolor}$a_{\h}$}};
\node at (1.732*\r,0.4*\r) {{\color{\acolor}$\k$}};
\filldraw[draw=\acolor,fill=\acolor] (1.732*\r,0) circle[radius=0.07*\r];
\draw[\acolor,line width=2 pt] (0.5*1.732*\r,0.5*\r) -- (0.5*1.732*\r,-0.5*\r);
\end{tikzpicture}
\end{align}
where the orientation of $p$ matches $e$ and the orientation of $p$ is opposite to $e$. 
The edge terms energetically enforce that the sector of a string matches the boundary label of the surrounding $G$ domains. 
The ground states of the symmetry-enriched string-net model correspond to superpositions of $G$-domains with consistent nets of $\mathcal{C}_G$ strings on the boundaries of the domains. 
This picture reveals a $G$ symmetry of the Hamiltonian and ground space, represented by a product of left group multiplications $\prod_p L_p(\g)$, where $L(\g)\ket{\h}=\ket{\mathbf{gh}}$. 
When applied to a region $R$ this symmetry fluctuates the domains by $\g$ and generates a domain wall described by $\mathcal{C}_{\g}$ at the boundary $\partial R$. 
The twist defects obtained by truncating the $\g$ domain walls are described by objects of the relative center $\mathcal{Z}_{\mathcal{C}_1}(\mathcal{C}_{\g})$~\cite{williamson2017}. 
The full emergent $G$-enriched topological order is described by the relative center $\mathcal{Z}_{\mathcal{C}_1}(\mathcal{C}_{G})$~\cite{gelaki2009centers}, which was denoted $\mathcal{Z}_{G}(\mathcal{C}_{G})$ in Ref.~\cite{williamson2017}.
Similar to the conventional string-net model, see Eqs.~\eqref{eq_appendix_half_braiding1} and~\eqref{eq_appendix_half_braiding2}, there are $G$-crossed half braidings which describe string operators for the anyons and defects of the emergent $G$-enriched topological order. 
In appendix~\ref{app_G_crossed} we describe solutions to the $G$-crossed half braiding equations that can be derived from a $G$-crossed modular braiding on the input category. 
These solutions are used in Eqs.~\eqref{eq_half_braiding1},~\eqref{eq_half_braiding2}.

The symmetry-enriched string-net model for $\mathcal{C}_G$ is phase equivalent to the conventional string-net model for $\mathcal{C}_1$. 
The equivalence is implemented by first adding $\mathbb{C}[G]$ degrees of freedom in the $\ket{+}$ state to the plaquettes of the $\mathcal{C}_1$ string-net, and extending the Hilbert space on each edge from $\mathbb{C}[\mathcal{C}]$ to $\mathbb{C}[\mathcal{C}_G]$.
Here the $\ket{+}$ state is defined by
\begin{align}
    \ket{+} = \frac{1}{\sqrt{|G|}} \sum_{\g\in G} \ket{\g}. 
\end{align}
Next, a local unitary circuit consisting of a product of controlled-$B_p^{\g}$ operators, $U=\prod_p CB_p$, is applied to the string-net Hamiltonian for $\mathcal{C}_1$, written with additional local terms that embedded the model into the larger Hilbert space for $\mathcal{C}_G$ described above. 
This local unitary circuit acts as an SET entangler, mapping the string-net Hamiltonian $H^{\mathcal{C}_1}_{\mathrm{SN}}$ to the symmetry-enriched string-net Hamiltonian $H_{G\mathrm{SN}}^{\mathcal{C}_G}$ as follows 
\begin{align}
    U \Big( H^{\mathcal{C}_1}_{\mathrm{SN}} -\sum_p \frac{1}{|G|} \sum_{\g} R_p(\g)  - \sum_e \Pi_e \Big) U^\dagger 
    \cong 
    H_{G\mathrm{SN}}^{\mathcal{C}_G}
    ,
\end{align}
where the additional plaquette terms project onto the $\ket{+}$ state, and the edge terms $\Pi \ket{a_{\g}} = \delta_{\g} \ket{a_{\g}}$ project onto the $\mathbb{C}[\mathcal{C}_1]$ string subspace. 
The controlled plaquette operators are defined by
\begin{align}
    CB_p\ket{\g }_p\ket{\psi}=\ket{\g}_p B_p^{\g}\ket{\psi}.
    \label{eq_def_controlled_plaquette_app}
\end{align}
Here, we extend the action of $B_p^{\g}$ to act as identity on states outside the support of $B_p^{1}$ such that $C B_p$ is a well defined local unitary operator. 
The equivalence relation above captures gap-preserving redefinitions of the Hamiltonian that preserve the ground space. 

Gauging the global $G$ symmetry of the symmetry-enriched string-net model can be achieved simply in this case by projecting each plaquette onto the $\bra{+}$ state~\cite{williamson2017}. 
This effectively removes the domain $G$ variables and allows the edge $\mathcal{C}_G$ string-nets to fluctuate freely and condense, recovering the conventional string-net model for $\mathcal{C}_G$. 
On the relevant Hamiltonians this is written
\begin{align}
    \bra{+}_P H_{G\mathrm{SN}}^{\mathcal{C}_G} \ket{+}_P \cong H_{\mathrm{SN}}^{\mathcal{C}_G},
\end{align}
where $\ket{+}_P = \bigotimes_p \ket{+}_p$. 
%----------
\section{Gauging symmetry-enriched string-nets with $G$-crossed modular input theories}
\label{app_G_crossed}
In this section, we review the derivation of the $G$-crossed theory $\mathcal{Z}_G(\mathcal{C}_G)$ that was presented in Ref.~\cite{williamson2017}. 
We describe how gauging realizes projectors onto different anyon types in the theory $\mathcal{Z}(\mathcal{C}_G)$.
For simplicity, we focus on a $G$ symmetry-enriched string-net with a $G$-crossed modular input category ${\mathcal{C}_{G}=\bigoplus_{\g}\mathcal{C}_{\g}}$, where $\g\in G$. 
We use the diagrammatic formalism for $G$-crossed braided fusion categories that was introduced in Ref.~\cite{PhysRevB.100.115147}, where a defect world line for $a_g\in\mathcal{C}_g$ is attached to a $g$-domain wall world sheet that extends back into the page.

We start from a modular tensor category $\mathcal{C}_1$. 
The anyonic excitations in the string-net based on $\mathcal{C}_1$ correspond to the double \begin{align}
\mathcal{Z}(\mathcal{C}_{\mathbf{1}})=\mathcal{C}_{\mathbf{1}}\boxtimes \mathcal{C}_{\mathbf{1}}^\mathrm{rev},
\end{align} 
where $\mathcal{C}_{\mathbf{1}}^\mathrm{rev}$ is the reverse (or opposite) category of $\mathcal{C}_{\mathbf{1}}$. 
This category is obtained from $\mathcal{C}_{\mathbf{1}}$ by reversing its braided crossings. 
Next, we introduce $G$-graded defects, which are described by $\mathcal{C}_G=\oplus_{\g\in G}\mathcal{C}_{\g}$. In particular, the initial category $\mathcal{C}_{\mathbf{1}}$ is the trivially graded sector with $\g=\mathbf{1}$.
Following the convention of Ref.~\cite{PhysRevB.100.115147}, we let the defects extends below the string-net lattice (into the page). 
The $G$-defects are introduced to the string-net model via the application of $\prod_p(CB)_p$ (see Eqs.~\eqref{eq_KW_EP_G} and \eqref{eq_CBg_psi}), the emergent theory becomes the $G$-crossed theory described by the $G$-\emph{relative center}
\begin{align}
    \mathcal{Z}_G(\mathcal{C}_G)=\mathcal{C}_{\mathbf{1}}\boxtimes \mathcal{C}_G^\mathrm{rev}.
\end{align}
We describe the emergent anyons and defects in this symmetry-enriched topological order, and their string operators, below. 

\subsection{Defect tube algebra}

The symmetry-enriched string-net lattice model, realizes $G$-enriched topological order. 
The anyons and twist defects in this symmetry-enriched topological order are defined by the corresponding \emph{defect tube algebra}~\cite{williamson2017}.
Let  $\Sigma$ be an annulus with two radii $0<r_1<r_2$,
\begin{align}
\Sigma=\{v\in \mathbb{R}^2: r_1\leq \norm{v}_2\leq r_2\}.
\end{align}
One can think of the tube or the annulus $\Sigma\subset \mathbb{R}^3$ living in the $x$-$y$ plane with the $z$-coordinates being zero.
Now we imagine thickening the annulus along the $z$-axis by multiplying it with $[-1,1]$ to clearly separate space above and below the $x$-$y$ plane. In particular,  the inner and the outer boundary circles are now both thickened to a cylinder $S^1\times [-1,1]$ (shown in gray below).
\begin{equation}
\vcenter{\hbox{\includegraphics[page=27]{Figures.pdf}}}
\end{equation}
One can view the system as living on planes parallel to the $x$-$y$ plane.  
On the bottom plane $\mathbb{R}^2\times \{-1\}$ live the defect strings $a_{\g}\in \mathcal{C}_{\g}^\mathrm{rev}$; on the top plane live the objects $a\in \mathcal{C}_{\mathbf{1}}$,
while the middle plane $\mathbb{R}^2\times \{0\}$ supports the original string-net, corresponding to the category $\mathcal{C}_1$, as well as the symmetry-enriched domain walls, with each labelled by  $\g\in G$. 
Following the convention from Ref.~\cite{PhysRevB.100.115147}, each domain wall terminates on a $\mathcal{C}_G$ string in the middle plane $\mathbb{R}^2\times \{0\}$ and extends back to $-\infty$ in the $z$-direction. Therefore, these domain walls act only on the bottom plane and do not interact with objects on the top plane.
Using the thickened tube picture, we now define a basis of defect tubes that correspond to emergent defects, and symmetry actions on these defects. 
For $\underline{a}_{\g}=(a,a_{\g})\in \mathcal{C}_{\mathbf{1}}\boxtimes \mathcal{C}_{\g}^\mathrm{rev}$, and $\g \in G$ the basis element is given by (here we have hidden the outer cylinder for clarity)
\begin{align}
    \B^{\h}_{\underline{a}_{\g}}:=\ \ 
   \vcenter{\hbox{\includegraphics[page=28]{Figures.pdf}}}
    ,\label{eq_B_h_ag}
\end{align}
where $a\in\mathcal{C}_1,a_{\g}\in \mathcal{C}_{\g}^\mathrm{rev}, ^{\h}a_{ \g}\in \mathcal{C}_{^{\h}\g}^\mathrm{rev}$, and  the dashed loop $\omega_{\h}$ around the inner cylinder in gray, representing an $\h$-domain wall, is a weighted sum 
\begin{align}
    \omega_{\h}:=\sum_{ a_{\h}\in \mathcal{C}_{\h}} \frac{d_{a_{\h}}}{\mathcal{D}_{\h}^2}\ \ 
   \vcenter{\hbox{\includegraphics[page=29]{Figures.pdf}}}
    .
\end{align}
The $\omega_\h$ domain walls only act on the bottom plane, therefore the defect line $a_{\g}$, after passing through the domain wall, is mapped to another defect line $^{\h}a_{\g}$ in a conjugated graded sector. 
The multiplication of the defect tube algebra is defined to be the composition of tubes and they form
a projective representation of the grading group $G$ \cite{williamson2017}
\begin{align}
  \B^{\h}_{b_{\f}}  \B^{\mathbf{k}}_{a_{\g}}=
  \delta_{b_{\f},^{\k}(a_{\g})}  
  \eta_a(\h,\k)\B_{a_{\g}}^{\h \mathbf{k}}.
\end{align}
Eq.~\eqref{B_op_projective_rep} is a special case of the equation above.
The defect tube algebra elements defined in Eq.~\eqref{eq_B_h_ag} generalize the tube algebra elements for the case of a modular tensor category (see for example, Eq.~(72) of \cite{KOENIG20102707}).

Recall that in the context of a string-net model, a Kirby $\omega_0$-loop is defined by the plaquette operator $B_p$, and it allows one to slide a string operator across each non-excited plaquette freely,  effectively removing the puncture in each plaquette in the string-net model.  The $\omega_{\h}$ loop defined here is a direct generalization of the Kirby loop~\cite{williamson2017}. 
It obeys the following property
\begin{align}
 \vcenter{\hbox{\includegraphics[page=30]{Figures.pdf}}}
  =
 \vcenter{\hbox{\includegraphics[page=31]{Figures.pdf}}}
 ,
  \end{align}
  for any $\h$ and any $\underline{a}_\g\in \mathcal{C}_{\mathbf{1}}\boxtimes\mathcal{C}_{\g}^\mathrm{rev}$.
%-----
Using this sliding rule, one can obtain the braiding structure of the objects $\mathcal{C}\boxtimes \mathcal{C}_G^\mathrm{rev}$ via the following procedure corresponding to a right handed braid of two punctures. To make this clear, we use a top-down perspective. 
As a result, the double-line shown in Eq.~\eqref{eq_B_h_ag} is drawn separately to avoid overcrowding the illustration, and  the thickened-puncture (i.e. the gray cylinder in Eq.~\eqref{eq_B_h_ag}) along with the $\omega_{\h}$-loop are now denoted collectively by a gray disk.
In particular, the  string $b$ on the top plane passes through the puncture with the loop $\omega_{\f}$ in a transparent way as it is not acted upon by the group element $\f$, while a $b_{\k}$ string on the bottom plane passes through the puncture and sends the $\omega_{\f}$ loop to $\omega_{\f \k}$. 
After the braiding $b$ is beneath $a$, while $b_{\k}$ is above $a_{\g}$, where the order of the crossings determines that the $b$ and $b_{\k}$ strings can be fused into the string-net lattice first, 
\begin{align}
\vcenter{\hbox{\includegraphics[page=32]{Figures.pdf}}}
=&
\vcenter{\hbox{\includegraphics[page=33]{Figures.pdf}}} ,
\label{eq:GxBAbove}
\\
\vcenter{\hbox{\includegraphics[page=34]{Figures.pdf}}}
=&
\vcenter{\hbox{\includegraphics[page=35]{Figures.pdf}}} .
\label{eq:GxBBelow}
\end{align}
Above, the string braidings are interpreted as being in the $G$-crossed modular theory $\mathcal{C}_{G}$. 
The strings above the plane reproduce the modular $\mathcal{C}_{\mathbf{1}}$ theory, while the strings below the plane reproduce the $G$-crossed modular theory $\mathcal{C}_G^{\mathrm{rev}}$, due to the reversed crossing with respect to the right handed exchange of the grey disks. 
Here, we reiterate that the strings above and below the $x$-$y$ plane do not interact, and hence a crossing between a string above the lattice with a string below is trivial. 
%--------

\subsection{Defect string operators}

 \begin{figure}[t]
\includegraphics[page=10]{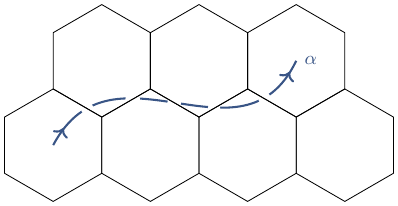}
\caption{Here we illustrate a string operator for anyon $\alpha$ in a symmetry-enriched string-net model.}
\label{string_op}
\end{figure}

One can resolve the $\mathcal{C}_G$ string diagrams shown above in Eqs.~\eqref{eq:GxBAbove},~\eqref{eq:GxBBelow}, and Figure~\ref{string_op}, into the string-net lattice using the $G$-crossed braiding theory of $\mathcal{C}_G$ to find explicit lattice operators. 
We refer interested readers to Section~6.5 of Ref.~\cite{KOENIG20102707} for a related discussion for modular input categories. 

We focus on objects in the emergent $G$-crossed theory that can be pictured under the lattice $\underline{\alpha}=(\mathbf{1},\alpha)\in \C_{\mathbf{1}}\boxtimes \C^\text{rev}_{G}$ where $\alpha$ is contained in $\C^\text{rev}_{G}$. 
Hence, we only need to draw the $\alpha$-string under the lattice.
The treatment of objects from $\mathcal{C}_{\mathbf{1}}$ above the lattice is similar to Ref.~\cite{KOENIG20102707}. 
For general input categories, string operators from the relative center $\mathcal{Z}_{\mathcal{C}_{\mathbf{1}}}(\mathcal{C}_G)$ can be found by solving $G$-crossed half-braiding equations~\cite{PhysRevB.103.195155} that generalize Eqs.~\eqref{eq_appendix_half_braiding1},~\eqref{eq_appendix_half_braiding2}. 
Here, we can directly resolve the $\alpha$-string into the lattice using the $G$-crossed modular braiding from the input category $\mathcal{C}_G$. 
This can be used to find $\Omega$-symbols that solve the $G$-crossed half-braiding equations for $\mathcal{Z}_{\mathcal{C}_{\mathbf{1}}}(\mathcal{C}_{G})$, which extend Eqs.~\eqref{eq_appendix_half_braiding1} and~\eqref{eq_appendix_half_braiding2} to the setting of symmetry-enriched string-net models,  
\begin{align}
\vcenter{\hbox{\includegraphics[page=11]{Figures.pdf}}}
=&\sum_{b,s,r}\Omega_\alpha^{a,rsb}
\ \ \sqrt{\frac{d_b}{d_a\sqrt{d_rd_s}}}\ \ 
\vcenter{\hbox{\includegraphics[page=12]{Figures.pdf}}} ,
\label{eq_half_braiding1}
\\
%-----------------
\vcenter{\hbox{\includegraphics[page=13]{Figures.pdf}}}
=&\sum_{b,s,r}(\Omega_\alpha^{a,srb})^*
\sqrt{\frac{d_b}{d_a\sqrt{d_rd_s}}}\ 
\vcenter{\hbox{\includegraphics[page=14]{Figures.pdf}}}
\, .
\label{eq_half_braiding2}
\end{align} 
After being resolved into the lattice, the string operator results in a superposition of labels $r,s\in \mathcal{C}_{\mathbf{1}}$ with some phase factors. 
The labels $r$ and $s$ account for the violation of fusion rules in the string-net of $\mathcal{C}_{\mathbf{1}}$, while the phase factor accounts for a generalized charge in the anyon $\alpha$. 

\subsection{Gauging defect projectors and string operators}

We implement gauging by measuring and correcting the plaquette degrees of freedom to $\bra{+}_P$, see Eq.~\eqref{eq_KW_EP_G}, the resulting theory is 
\begin{align} 
\mathcal{Z}(\mathcal{C}_G)=\mathcal{C}_{\mathbf{1}}\boxtimes (\mathcal{C}_G^{\mathrm{rev}}//G),\label{eq_Z_C_G}
\end{align}
where $\mathcal{C}_G^{\mathrm{rev}}//G$ denotes a $G$-\emph{equivariant} theory.
We now discuss the anyons in the $G$-equivariant theory. 
These anyons arise from irreducible representations on orbits of anyons and defects in the $G$-crossed modular theory.
We define the \emph{orbit} of a defect under the $G$-action to be
\begin{align}
     [a_{\g}]:=\{\rho_{\h}(a_{\g})| \h\in G\},
\end{align}
where we have used the notation of Ref.~\cite{PhysRevB.100.115147}.
This corresponds to the set of defects that can be reached by acting on $a_{\g}$ via symmetries. 
This data is given as part of a $G$-crossed braided category. 
Choosing $a_{\g}\in [a_{\g}]$ as a representative of the orbit, we define its \emph{centralizer} as the elements invariant under the group action
\begin{align}
     Z(a_{\g}):=\{\h\in G|\rho_{\h}(a_{\g})=a_{\g}\}.
\end{align}
A simple object in $\mathcal{Z}(\mathcal{C}_G)$ is given by a pair $(c,\alpha)$ with $c\in \mathcal{C}_{\mathbf{1}}$ and $\alpha\in \mathcal{C}_G^\mathrm{rev}//G$. Each simple object $\alpha\in \mathcal{C}_G^\mathrm{rev}//G$ consists of a pair $\alpha=([a_{\g}],R)$ with $R$ a projective representation of $Z(a_{\g})$.  Each simple object $\alpha$ in $ \mathcal{C}^\mathrm{rev}//G$ corresponds to an irreducible idempotent 
\begin{align}
     \Pi_{\alpha}= \sum_{\h\in Z(a_{\g})}\chi_R^*(\h)\B_{a_{\g}}^\h.
\end{align}

Recall that the gauging procedure described for ground states in the main text involves a measurement $\bra{+}_P$, which gives projections to anyons that have trivial gauge charge. 
The gauging procedure described here, and for the defect tube algebra in Section~\ref{subsec:GaugingAnyonTubes}, follow Ref.~\cite{williamson2017} and can be applied to states containing anyon and defect excitations. 
When combined with the defect tube irreducible central idempotent solutions in Eq.~\eqref{eq_B_h_ag}, and the string operators in Eqs.~\eqref{eq_half_braiding1},~\eqref{eq_half_braiding2}, we arrive at a method to derive the anyons and string operators of a string-net with a $G$-crossed modular input category via gauging. 
This generalizes the recipe to construct string-operators and irreducible central tube idempotents for string-net models with a modular tensor category input, as explained for example in Ref.~\cite{KOENIG20102707}.
%----------

\section{Quantum circuits for $CB_p$ operators}
\label{app:CBCircuit}

In this appendix, we present an explicit circuit realization of the controlled-plaquette operator for the Ising category. 
For readers with a quantum-computational focus, the following section is intended to facilitate a detailed understanding of the Levin-Wen string-net model. In particular it is inteded to describe what the plaquette operator $CB_p$ corresponds to as a quantum circuit. 
Readers who are not interested in such details may skip the remainder of this subsection and simply accept that $CB_p$ is in general constructible via a quantum circuit. 

To detail the construction of $CB_p$, we first introduce the $F$-move or $F$-symbol.  
We use the convention  below.
\begin{align}
\vcenter{\hbox{\includegraphics[page=1]{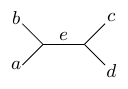}}}
=\sum_f F^{abc}_{def}
\vcenter{\hbox{\includegraphics[page=2]{Figures.pdf}}}
\end{align}
The $F$-move is essentially a 4-on-1 controlled unitary. 
All the $F$-symbols for Ising are given by Table~1 in Ref.~\cite{Kitaev_2006}. Namely the labels $\{a,b,c,d\}$ correspond to the controls while $e$ and $f$ correspond to the single target qudit. 
We henceforth abuse notation to use $\{a,b,c,d\}$ and $e$ to indicate the control qudits and the target qudit respectively, separated by a semicolon, as in $\ket{abcd;e}$. We do not need $f$ since $e$ and $f$ correspond to the same qubit. 
Hence we write
\begin{align}
    F=CU_{abcd\rightarrow e}\label{eq_F_symbol} ,
\end{align}
for example for fixed control $a=b=c=d=\sigma \in \text{Ising}$,
\begin{align}
    F\ket{\sigma\sigma\sigma\sigma;e}
    =\ket{\sigma\sigma\sigma\sigma}\frac{1}{\sqrt{2}}\mqty(1& 1 \\1 &-1)\ket{e}, \label{eq_F_sigma_e}
\end{align}
for $e\in\{\mathbf{1},\psi\}$.
We let the 2-by-2 unitary in Eq.~\eqref{eq_F_sigma_e} be denoted ${M\equiv [F^{\sigma\sigma\sigma}_\sigma]}$.
To promote this into a unitary on the qutrit space spanned by states $\{\mathbf{1},\psi,\sigma\}$, one can extend  the 2-by-2 matrix $M$ into a 3-by-3 unitary. The choice of how one extends $M$ does not matter since we are always in the string-net subspace, and hence $\ket{e}$ can only be $\mathbf{1}$ or $\psi$ when the four qudits labeled by $\{a,b,c,d\}$ are all $\sigma$. One can similarly extend the action of the F-symbol in other blocks, characterized by the four control qudits. 
With the unitary $F$ gate defined above, we now define a controlled-$F$ gate with the control being the qubit corresponding to the grading of $\sigma$,
\begin{align}
    C_pF:=\ket{0}\bra{0}\otimes \mathds{1}+\ket{1}\bra{1}\otimes F,
\end{align}
which is a 6-qubit gate, with a qubit at the center of plaquette $p$ and the others acted on by the target $F$-gate. 
To summarize, $C_pF$  is essentially a controlled unitary with five controls and one target.

\begin{figure*}
\begin{equation*}
 %--------------
 % the first hexagon
 %--------------
 \vcenter{\hbox{\includegraphics[page=3]{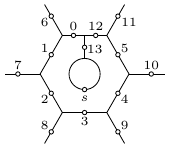}}}
\ \to \ 
%--------------
 % the second hexagon
 %--------------
    \vcenter{\hbox{\includegraphics[page=4]{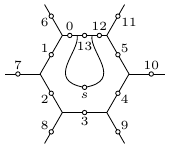}}}
\  \to \ 
%--------------
 % the 3rd hexagon
 %--------------
    \vcenter{\hbox{\includegraphics[page=5]{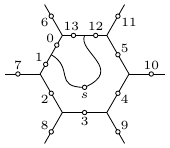}}}
 \ \to \ \cdots \ \to \ 
 %--------------
 % the 4th hexagon
 %--------------
    \vcenter{\hbox{\includegraphics[page=6]{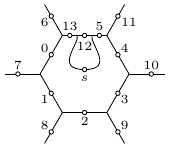}}}
 \end{equation*}
 \caption{The consecutive steps to apply the plaquette operator $B_p^\sigma$ for the doubled Ising topological order.}
 \label{consecutive_steps_of_CB}
 \end{figure*}

In fact, for our purpose here $c$ is always $\sigma$, and hence we can omit the control qutrit $c$. Inspired by the construction in Ref.~[55], we list below the steps of a procedure to implement the plaquette operator using a quantum circuit, which is illustrated in Figure~\ref{consecutive_steps_of_CB}. 
\begin{itemize}
    \item[(1)]  Introduce three ancilla qutrits labeled by $q_{12},q_{13},q_s$ initialized in $\ket{\mathbf{1}},\ket{\mathbf{1}},\ket{\sigma}$ respectively. 
    \item[(2)] Let $U_{0,12}$ be a unitary  such that
    \begin{align}
       U_{0,12} :\ket{i}_0\ket{\mathbf{1}}_{12}\mapsto \ket{i}_0\ket{i}_{12},
    \end{align}
    for $i\in\mathcal{C}^{(1)}\subset\mathcal{C}^{(2)}$ where $\mathcal{C}^{(1)}=\{\mathbf{1},\psi\}$ and $\mathcal{C}^{(2)}=\{\mathbf{1},\psi,\sigma\}$. Note that $U_{0,12}$ behaves like a $\operatorname{CNOT}$ gate when restricted to the subspace $\mathcal{C}^{(1)}$ for each qudit. Perform this unitary to the circuit.
    \item[(3)] Apply $CF$ with label assignment $a\to 0,b\to 12,c\to s$, $d\to s$ (omitted) and $e=13$. After this step the ancilla $7$ can be viewed as a horizontal $\sigma$.
    \item[(4)] Apply $CF$ with $a\to 1,b\to 6,c\to 13,d\to s,e\to 0$. This moves $s$ and $0$ to the topleft edge (previously labeled by 1).  Therefore there are two edge qutrits (0 and 1) on the topleft edge with the $s$-curve connecting them.
    \item[(5)] Repeat the previous step for each vertex of the plaquette.  
    \item[(6)] One ends up with three qutrits on the top edge, labeled by $13,12$ and $5$. Remove ancilla $12,13$ and $s$. 
\end{itemize}

%----------
 
%--------
The controlled-plaquette operator for Ising has the property that
\begin{align}
(CB_p)^2=\operatorname{id}\otimes \frac{1}{2}(B_p^{\mathbf{1}}+B_p^{\psi}).
\end{align}
This implies that the controlled operator $CB_p$ is only invertible in the subspace of 
\begin{align}
    B_p^{0}\equiv\frac{1}{2}( B_p^{\mathbf{1}}+B_p^{\psi})=+1.
\end{align}
at plaquette $p$.
In the total Hilbert space, we have to extend such an operator $CB_p$ by inserting something in the orthogonal subspace  $\mathds{1}-B_p^0$. One natural  way is to simply insert some unitary $U$. Therefore
\begin{align}
    \overline{CB_p}:=\mqty(CB_p& 0 \\ 0& U),\label{eq_extended_CB}
\end{align}
where we have abused the notation $CB_p^\sigma$ in the block matrix form to indicate its restriction to its support subspace $B^{0}_p$. Note that the choice of $U$ does not matter as after the first step of gauging we are in the subspace of $B_p^{0}=+1$, i.e.~the top left block in Eq.~\eqref{eq_extended_CB}.

The procedure shown Fig.~\ref{consecutive_steps_of_CB} was described for pedagogical illustration. 
In a practical experimental setup, one can remove the ancilla $s$ as it is in a fixed state, thereby reducing the number of qudits in use. 
One can also perform six controlled-$F$ on the six vertices in parallel to save operational time. 
See, for example, the procedure illustrated in Eq.~(C1) of Ref.~\cite{Levin_2005}.

The procedure described above is not unique to the Ising string-net example and can be extended to other graded categories by using the appropriate $F$-symbol, and grading, data. 
This generalization requires the implementation of controlled-$B_p^\g$ operators, which may involve a sum over more than one $B_p^{s_\g}$ operator, see Eq.~\eqref{eq_B_p_grading_sectors_appendix}. 
We focus our attention on states that are in the $+1$-eigenspace of $B_p^{\mathbf{1}}$, as these are the states we aim to gauge, in the absence of anyonic excitations. 
The following equation,  
\begin{align}
    B^{s_{\g}}_p B^{\mathbf{1}}_p=d_{s_{\g}} B^{\g}_p,
    \label{eq_representing_g}
\end{align}
then implies that the desired operator $B^{\g}_p$ and the simpler operator $\frac{1}{d_{s_{\g}}}B^{s_{\g}}_p$ act identically within the relevant $+1$-eigenspace of $B_p^{\mathbf{1}}$. 
Hence, we can implement the desired controlled-$B_p^\g$ plaquette operator on the relevant subspace by instead implementing a simpler controlled-$\B^{s_{\g}}_p$ operator, as described above and in Figure~\ref{consecutive_steps_of_CB}. 
Applying this to a string-net ground state $\ket{\Psi}$ and an ancilla in a superposition state over group elements we find 
\begin{align}
(CB^{s_\g})_p\sum_{\g}\frac{1}{d_{s_{\g}}}\ket{\g}_p \ket{\Psi}
&=
\sum_{\g}\qty(\frac{B_p^{s_{\g}}}{d_{s_{\g}}})\ket{\g}_p\ket{\Psi}
\nonumber \\
&=\sum_{\g}(B^{\g})_p\ket{\g}_p\ket{\Psi}
\nonumber \\
&= (CB^{\g})_p\sum_{\g}\ket{\g}_p\ket{\Psi} ,
\end{align}
where we have used that $B_p^{\mathbf{1}} \ket{\Psi} = \ket{\Psi}.$
The ancilla qudit can then be measured in the irreducible representation basis to gauge the plaquette. This results in a projection onto a definite abelian anyon charge, which can then be removed following the procedure described in the main text. 

In the presence of nontrivial anyons or defects, a similar strategy to implement the plauqette operators required for gauging applies. This follows from a generalization of Eq.~\eqref{eq_representing_g} to the action of an $s_\g$ string on $\B^{\mathsf{1}}_{\underline{a}_{\h}}$, the projector onto an anyon or defect $a_{\h}$, and the domain wall operator $\B^{\g}_{\underline{a}_{\h}}$, see Ref.~\cite{williamson2017}.

\section{Background on $TY(\mathbb{Z}_3)$ and $SU(2)_4$}
\label{app:TY}

In this appendix we present background material about the fusion categories $TY(\mathbb{Z}_3)$ and $SU(2)_4$, see Refs.~\cite{Huston2022,Vanhove_2022} for related discussions.
In $TY(\mathbb Z_3)$, the simple objects are labelled $0,1,2,\sigma$ with $\mathbb{Z}_3$ fusion rules on the $0,1,2$ subtheory and 
\begin{align} 
\sigma \otimes \sigma = 0 \oplus 1 \oplus 2 .
\end{align}
Note that 
$\Vec_{\mathbb Z_3} = 0 \oplus 1\oplus 2$ 
can be given a non-degenerate braiding. There are two possible choices. One corresponds to $T= \text{diag}\{1,e^{2\pi i/3},e^{2\pi i/3}\}$, and the other is the time-reversed version. Following the notation of Ref.~\cite{BondersonThesis} we call these two choices $\mathbb Z_3^{(1)}$ and $\mathbb Z_3^{(-1)}$. 

We consider the first choice for now. This particular anyon theory can be considered a subgroup of anyons in the $D(\mathbb Z_3)$ TC as 
\begin{align}
    \mathbb{Z}_3^{(1)} = \{1,e m,  e^*m^*\}, \quad \mathbb{Z}_3^{(-1)} = \{1,e^* m,  em^* \}.
\end{align}
Hence, $D(\mathbb Z_3) = \mathbb Z_3^{(1)} \boxtimes \mathbb Z_3^{(-1)}$.

The center of $TY(\mathbb Z_3)$ is obtained by gauging the $e-m$ duality symmetry of $D(\mathbb Z_3)$, which keeps $ \mathbb Z_3^{(1)}$ invariant but acts non-trivially on $ \mathbb Z_3^{(-1)}$.
The mapping to the center of the gauged theory is labelled by the orbit of the action of the grading group and the representation of the centralizer group $Z_a$ of the orbit \cite{williamson2017}, where $a$ is a representative of the orbit.  There are four objects in the defect SET, partitioned into three orbits: 
\begin{align}
    O_1:=\{1\},\quad  O_{em}:=\{em^*,e^*m\}, \quad O_\sigma:=\{\sigma\}.
\end{align}
%---------
 The centralizer groups  $Z_1$ and $Z_\sigma$ are the grading group $\mathbb{Z}_2$ itself, and therefore they are doubled by the representations of the centralizer group (see Eq.~(300) of Ref.~\cite{williamson2017})
\begin{align}
    \mathbf{1}=(O_1, +),\quad z=(O_1,-),\quad \sigma_\pm:=(O_\sigma,\pm).
\end{align}
where $+$ ($-$) represents the trivial (sign) representation of $\mathbb{Z}_2$. The centralizer group $Z_{em}$ for the orbit $\{em^*,e^*m\}$ is trivial (denoted by $+$), therefore 
\begin{align}
    \phi=(O_{em},+).
\end{align}
The fusion rules are 
\begin{align}
    &\phi \otimes \phi = 1 \oplus z \oplus \phi, \\
    &\sigma_\pm \otimes \sigma_\pm = 1\oplus \phi, \quad 
    z\otimes \sigma_\pm= \sigma_\mp\\
&\sigma_\pm \otimes \phi=\sigma_+  \oplus \sigma_-, \quad \phi\otimes z=\phi.
\end{align}
Together, 
\begin{align}
\{1, \sigma_+, \phi, \sigma_-, z\} \equiv \{0,\frac{1}{2},1,\frac{3}{2},2 \}  ,
\end{align}
in $\overline{SU(2)_4}$ Chern-Simons theory. 
There is also a $\mathbb{Z}_3^{(1)}$ factor in the topological order, which corresponds to the strings $\{\mathbf{1},em,e^*m^*\}$ over the lattice (objects in $\mathcal{C}_{\mathbf{1}}$ that are introduced on top of the string-net plane $\mathbb{R}^2\times \{1\}$). All of the string operators can be encoded in the half-braiding data of the original topological order $D(\mathbb{Z}_3)$, with $\Omega^{a,rr(r+a)}=\omega^{ar}$. 
We use the convention that the $G$-crossed braiding of an $r$-string over the string-net plane is
\begin{align}
    \Omega^{\sigma,rr\sigma}=\omega^{r^2},
\end{align}
which agrees with the expectation that the $\sigma$-domain wall does not interact with objects on the top plane (that is, it does not flip $r$ to $-r$). 
Then an anyon in the center of $TY(\mathbb{Z}_3)$ can be represented as a pair
\begin{align}
    (a,b),\quad a\in \{\mathbf{1},em,e^*m^*\},\quad b\in \{\mathbf{1},\sigma_\pm, \phi,z\},
\end{align}
see Eq.~\eqref{eq_Z_C_G}. 
To conclude, $\mathcal{Z}(TY(\mathbb Z_3)) = \mathbb Z_3^{(1)} \boxtimes \overline{SU(2)_4}$.

We remark that similarly gauging the $e-m^*$ duality symmetry of $D(\mathbb Z_3)$ would instead give $ SU(2)_4 \boxtimes \mathbb Z_3^{(-1)}$. Finally, gauging both $\mathbb Z_2$ symmetries results in the double of $SU(2)_4$.

Above we gauged the $\mathbb Z_2$ symmetry assuming a trivial cocycle in $H^3(\mathbb Z_2,U(1))$. 
Choosing the non-trivial cocycle (i.e. stacking with the Levin-Gu SPT before gauging), we instead find $\overline{JK_4}$ which has the same fusion rules, but different F/R-symbols. 
We comment that there are fifteen minimal central idempotents corresponding to the simple objects, i.e.~anyons. 
In particular, 
\begin{align}
    P^{\mathbf{1}}=&\frac{1}{6}\sum_{r\in \mathbb{Z}_3}\mathcal{T}^{s}_{0s0}+\frac{\sqrt{3}}{6}\mathcal{T}^\sigma_{0\sigma 0}, \label{eq_P_1}\\
    P^{\Phi}=&
    \frac{1}{6}\sum_{r=1}^2\qty(\sum_{s,k\in \mathbb{Z}_3}\mathcal{T}^s_{r(r+k)r}\omega^{-rs}+\sqrt{3}\mathcal{T}^\sigma_{(-r)\sigma r}\omega^2),\label{eq_P_Phi}
\end{align}
where the two terms in $P^{\Phi}$ are the same as those in Eq.~\eqref{eq_c_spqr}.  This protocol can also be used to prepare anyons $\Phi_1=(em,\phi)$ and $\Phi_2=(e^*m^*,\phi)$ as fusing strings from the first layer commutes with the defectification and gauging process. 

In general, finding the irreducible central idempotents for the gauged anyons requires calculating the defect tube algebra, including irreducible central idempotents for defects, their symmetry actions, and the resulting projective irreducible representations. 
For the special case where a $G$-crossed modular category is input to the string-net, this process is significantly simplified; see Appendix~\ref{app_G_crossed}. 
%-------------
\end{document}